\documentclass[%
superscriptaddress,
twocolumn,
10pt,
nofootinbib,
 amsmath,amssymb,
 aps,
 pra,
floatfix]{revtex4-2}

\usepackage{graphicx,array}
\usepackage{dcolumn}% Align table columns on the decimal point
\usepackage{multirow}
\usepackage{array}
\newcolumntype{P}[1]{>{\centering\arraybackslash}p{#1}}
\usepackage{siunitx}
\usepackage{enumitem}

\usepackage{xcolor}

\definecolor{magenta}{rgb}{1.0,0.0,1.0}
\definecolor{green}{rgb}{0.0,0.8,0.4}
\definecolor{purple}{rgb}{0.5,0.0,0.5}

\usepackage{hyperref}% add hypertext capabilities
\usepackage{longtable}
\begin{document}
%%%%%%%%%%%%%%%%%%%%%%%%%%%%%%%%%%%%%%%%%%%%
\title{Network analysis of graduate program support structures\\ through experiences of various demographic groups}

\author{Robert P. Dalka}
\email{rpdalka@umd.edu}
\affiliation{University of Maryland, College Park, MD 20742, USA}

\author{Justyna P. Zwolak}
\email{jpzwolak@nist.gov}
\affiliation{National Institute of Standards and Technology, Gaithersburg, MD 20899, USA}
\affiliation{Joint Center for Quantum Information and Computer Science, University of Maryland, College Park, Maryland 20742, USA}
\date{\today}% 
%%%%%%%%%%%%%%%%%%%%%%%%%%%%%%%%%%%%%%%%%%%%%%%%%%%%%%%%%%%%%
\begin{abstract}
Physics graduate studies are substantial efforts on the part of individual students, departments, and institutions of higher education.
Understanding the factors that lead to student success and attrition is crucial for improving these programs.
One factor that has recently started to be investigated is the broadly defined students' experiences related to support structures.
The Aspects of Student Experience Scale (ASES), a Likert-style survey, was developed by researchers to do just that.
In this study, we leverage the network approach for Likert-style surveys (NALS) methodology to provide a unique interpretation of responses to the ASES instrument for well-defined demographic groups. % .
We confirm the validity of our findings by studying the stability of the NALS themes and investigating how they are expressed within demographic-based networks.
We find that for all four themes in the original ASES study, certain thematic trends capturing students' experiences vary across the demographic-based networks in meaningful ways.
We also reveal that for some demographic groups, there is an interesting interplay between, and mixing of, the original themes. 
Finally, our study showcases how NALS can be applied to other Likert-style datasets.
\end{abstract}
\maketitle

%%%%%%%%%%%%%%%%%%%%%%%%%%%%%%%%%%%%%%%%%%%%%%%%%%%%%%%%%%%%%
%---INTRODUCTION---%
%%%%%%%%%%%%%%%%%%%%%%%%%%%%%%%%%%%%%%%%%%%%%%%%%%%%%%%%%%%%%
\section{Introduction}\label{sec:introduction}
%%%%%%%%%%%%%%%%%%%%%%%%%%%%%%%%%%%%%%%%%%%%
Graduate studies is an intensive, high-resource endeavor, both for the individual student and for the institution.
While many physics programs aim to offer support for their students in completing the doctoral programs, a significant number of graduate students who enroll in a program do not obtain their degree~\cite{Lovitts00HCG, Sowell15-DIM}.
Investigating what leads to graduate attrition and the types of support structures that can lead to success is of high concern for researchers in education~\cite{Collier23-THU, OMeara17-SOB, Moreira19-IPD}.

To date, the majority of research efforts in the physics education research (PER) community have investigated undergraduate attrition and persistence.
Most of these studies focus on a variety of factors that can make up a student's experience, such as a sense of belonging~\cite{Lewis16-FOO, Whitcomb23-PSE, Dortch17-BUW}, physics identity~\cite{Hyater18-CLP, Hazari10-CHS, Close16-BPP}, self-efficacy~\cite{Dou16-BPM, Dou18-UDI}, and classroom interactions~\cite{Zwolak17-IIP, Zwolak18-PIN}.
This work has helped to make recommendations for departments on how to improve their physics undergraduate programs to support better student experiences~\cite{Beckford20-TIS, McKagan21-EP3}.

Graduate programs in physics have been a growing interest in PER.
Recent studies have investigated individual students and particular minoritized student experiences~\cite{Barthelemy20-ESG, Scherr17-FGM, Sachmpazidi23-REP, Renbarger23-DDA}, as well as changes in graduate programs~\cite{Barthelemy23-GPR, Posselt17-EEB}.
Additionally, the American Physical Society (APS) Bridge Programs has gathered best practices for creating a graduate program to recruit and support minoritized students in physics~\cite{bp-manual}.
However, the physics graduate student experience and other effects on student attrition remain largely unexplored.
Understanding how physics departments support students in their graduate studies is an important step in identifying how graduate students experience their programs.
The priorities of what support structures exist are reflections of the larger culture in physics departments.
Taking a systems approach can help identify how to effectively mitigate challenges and increase graduate student success.

To capture the experiences of support structures available to physics graduate students, researchers developed the Aspects of Student Experience Scale (ASES) instrument (referred to hereafter as ASES)~\cite{Sachmpazidi21-DSS}.
Like other Likert-style surveys, ASES includes items that ask respondents to indicate the level they agree or disagree with various statements.
These statements relate to different types of support structures that graduate students may experience and that would help them in their graduate program.
ASES was developed to align with the APS Bridge Program recommendations for student support, including support for social integration, academic success, positive mentorship, strong research experiences, building professional skills, and financial well-being~\cite{bp-keycomp}
The ASES instrument was administered to graduate students in physics departments across the United States.

In our previous work, we have proposed a network analysis approach to analyze Likert-style surveys, such as ASES, to reveal themes driven by student responses~\cite{Dalka22-NALS}.
Approaching a Likert-style survey as containing items that may be uniquely linked to each other aligns with a model of experiences of individual phenomena being linked to each other to form larger themes.
The network analysis for Likert-style surveys (NALS) methodology is designed to enable the modeling of survey items as a network built from survey responses. 
The individual survey items are the nodes, and the edges are weighted by the similarity in responses between those nodes.
NALS offers new opportunities for the interpretation of survey data.
Compared to other methods, such as exploratory factor analysis, NALS is unique in that along with capturing the high-level thematic clusters, it allows investigation of the multilevel complexities of subclusters within each theme.
We have also demonstrated how node-centric measures can help to identify important items within the network structure.
These types of investigations are made possible through NALS.

Using the set of responses to the ASES instrument (hereafter referred to as the ASES dataset), we show that the large-scale clusters of items revealed using NALS are informative about larger themes that exist in the survey.
However, one may also be interested in how the larger thematic groups of items are subject to the respondents who generated those groupings.
A grouping of survey items that captures the experiences of the full respondent population may not necessarily be representative of the experiences of certain demographic groups of students.
Our research questions focus on how the emerging themes revealed by NALS compare between different demographic groups and what these differences mean for student experiences. 
We also investigate the stability of the emerging themes.
A more explicit statement of these research questions is provided in Sec.~\ref{sec:results}, after adequate background information is described.

In this paper, we investigate this by comparing how the partitioning of survey items into thematic groups changes when the survey networks are built based on responses from well-defined demographic groups.
We also test the stability of the new themes against small changes in the sample populations to validate the underlying structure of the demographic-based networks.
We find that the NALS method, combined with clustering techniques and network centralities, allows us to identify meaningful features unique to each network.

The manuscript is organized as follows: In Sec.~\ref{sec:background}, we present theoretical grounding for the choices of demographic variables we consider in this work.
In Sec.~\ref{sec:method}, we further describe the ASES instrument, give a summary of the NALS methodology for creating networks, and introduce the network analysis techniques used in our work.
In Sec.~\ref{sec:results}, we present the results of the NALS analysis for both the full network and the demographic-based networks.
The interpretation of the results and discussion of what they mean for the ASES themes concerning different demographic groups are presented in Sec.~\ref{sec:inter}.
Finally, in Sec.~\ref{sec:conclusion}, we summarize the results and suggest future research directions.

%%%%%%%%%%%%%%%%%%%%%%%%%%%%%%%%%%%%%%%%%%%%%%%%%%%%%%%%%%%%%%
\section{Background}\label{sec:background}
%%%%%%%%%%%%%%%%%%%%%%%%%%%%%%%%%%%%%%%%%%%%%%%%%%%%%%%%%%%%%%
This paper makes contributions to the literature in two primary ways. 
First, we show how the use of NALS to analyze survey responses can provide insights when comparing groups of respondents. 
These insights are gained through the use of network analysis tools that are introduced in Sec.~\ref{ssec:net_analysis}. 
Second, we present findings specifically relevant to the ASES instrument and the different groups that were important for the development of this survey tool. 

The student groups chosen in this study were limited due to the number of responses---in order to obtain a stable network. 
As shown in Appendix~\ref{app:sample-size}, the smaller the respondent pool, the more unstable the network is. 
We chose groupings of respondents that we could reasonably hypothesize may have different experiences of graduate student support structures due to their comparable sample sizes. 
In this section, we present literature about how these different student groups experience physics and science, technology, engineering, and mathematic (STEM) graduate programs.

%%%%%%%%%%%%%%%%%%%%%%%%%%%%%%%%%%%%%%%%%%%%%%%%%%%%%%%%%%%%%%
\subsection{Bridge and nonbridge programs}
%%%%%%%%%%%%%%%%%%%%%%%%%%%%%%%%%%%%%%%%%%%%
The ASES instrument was designed as a tool for capturing the ways that physics graduate programs provide support for their students through their experiences~\cite{Sachmpazidi21-DSS}. 
The ASES items were designed to capture the preidentified support structures that the APS Bridge Program~\cite{apsbp} recommended for bridge sites to adapt and implement to better support bridge students in their programs~\cite{Hodapp17-BBU, Hodapp18-APS, Casas22-EFF}.
Bridge sites are expected to offer a range of support services for their students centered on the engagement and commitment of faculty, mentorship, multifaceted admissions practices, monitoring of students' progress, and recordkeeping of demographics~\cite{apsbp-partners}.
The ASES was designed to assess only the support structure that students could report on (excluding, for example, the department’s practices around recordkeeping on student demographics or admissions).
Over time, besides the bridge sites, several bridge partnership sites chose to adopt these practices and apply them across all bridge and nonbridge students.

In our analysis, we examine the network models of students in bridge-affiliated sites (bridge and bridge partnership sites) and nonbridge-affiliated sites.
Our goal is to investigate the kinds of differences NALS reveals for students with and without intentionally designed support structures.

%%%%%%%%%%%%%%%%%%%%%%%%%%%%%%%%%%%%%%%%%%%%%%%%%%%%%%%%%%%%%%
\subsection{Women and men}
%%%%%%%%%%%%%%%%%%%%%%%%%%%%%%%%%%%%%%%%%%%%
The underrepresentation of women in physics is a persistent and well-recognized issue across science and technology fields~\cite{Sax16-WIP, Kanny14-IFY, Ivie19-WIP}.
It is widely documented in the literature that the underrepresentation of women is a result of the social and cultural environments of science fields that lead to unwelcoming spaces for women~\cite{Barthelemy16-GDP}.
In these spaces, the resources and support within the programs will be better tailored for men and will not serve women students in the same ways~\cite{Cabay18-CCB, DeWelde11-GOC}.
Interactions with peers and research mentors are important for the persistence of women in STEM graduate programs as they foster a sense of belonging~\cite{Barthelemy20-ESG, Stockard21-EWU}.
These forms of support are well represented in the ASES instrument.
In the analysis presented in this paper, we aim to identify the ways in which experiences of different support structures are interconnected in unique ways for men and women physics graduate students.
In turn, we identify areas in which programs can better support women students through intentionally connecting particular program structures.

Nonbinary and other gender minority respondents were not included in this analysis as there were not enough of these responses to create a reliable network model. This is a limitation of this methodology but not a unique limitation among quantitative analysis.

%%%%%%%%%%%%%%%%%%%%%%%%%%%%%%%%%%%%%%%%%%%%%%%%%%%%%%%%%%%%%%
\subsection{Early and later semesters}
%%%%%%%%%%%%%%%%%%%%%%%%%%%%%%%%%%%%%%%%%%%%
Another factor that informs what support structures students may need is the stage of the graduate program they are currently at.
Graduate school involves many small steps along the way toward a doctoral degree.
This might involve the completion of classes, finding a research group, or reaching candidacy.
In the ASES dataset, while there are no demographic questions that specifically ask about those stages, respondents report how many semesters they have been a part of their program.
To approximate the stage of completing classes, we have grouped respondents into less than 2 years (four semesters) in their program and greater than 2 years.
This is in line with when most physics doctoral programs expect students to have completed course requirements~\cite{Goldberg-TBA}.
Additionally, this is when students often start applying for and transitioning to candidacy.
Thus, we can expect that the support that students in their ``early'' semesters will need will be different from the support that students in their ``later'' semesters will need.
We expect that these differences in need translate to differences in reported experiences.
In our study, we aim to identify how this affects the interconnectedness of the support structures as measured by the ASES instrument and modeled by NALS.

%%%%%%%%%%%%%%%%%%%%%%%%%%%%%%%%%%%%%%%%%%%%%%%%%%%%%%%%%%%%%%
\subsection{Funding sources} 
%%%%%%%%%%%%%%%%%%%%%%%%%%%%%%%%%%%%%%%%%%%%
Researchers have shown that across STEM doctoral programs, students funded through research assistantships are more likely to complete their degrees than their peers~\cite{Ampaw12-CTS}. 
This indicates that this form of financial support has implications for how students feel supported in their program through other experiences. 
Additionally, the skills that graduate students gain when they are funded through a research assistantship or a teaching assistantship will differ and are related to the type of work they are engaging in~\cite{Grote21-SDS}. 
The support that graduate students experience when engaging in research or teaching will be different. 
In the ASES dataset, there are students who have been funded through multiple sources. 
To capture the possibly distinct experiences due to funding sources, we thus have organized into three categories: research funding, nonresearch funding, and mixed funding. 
In understanding how different experiences of support structures are connected, or not connected, we can model how the experiences of students who have solely been supported by research funding compare to those who have relied on other sources of funding.

%%%%%%%%%%%%%%%%%%%%%%%%%%%%%%%%%%%%%%%%%%%%%%%%%%%%%%%%%%%%%%
\section{Methodology}\label{sec:method}
%%%%%%%%%%%%%%%%%%%%%%%%%%%%%%%%%%%%%%%%%%%%%%%%%%%%%%%%%%%%%%
\subsection{The ASES instrument}
\label{ssec:ases}
%%%%%%%%%%%%%%%%%%%%%%%%%%%%%%%%%%%%%%%%%%%%
The data for this study come from graduate student responses to the ASES instrument~\cite{Sachmpazidi21-DSS}.
ASES was designed to capture physics graduate student experiences with various support structures that may exist within graduate programs.
The survey items were developed in partnership with the American Physical Society Bridge Program~\cite{apsbp} and based on prior literature that shows what supports are important for complete educational experiences in bridge programs.
The dataset was collected in Spring 2019, with ASES being administered to graduate students within 20 physics departments (from the starting call to 60, with one department being dropped from the dataset due to low response rate)~\cite{Sachmpazidi22-SEP}.
Due to only using fully complete responses in our analysis, the original 397 responses are filtered down to 381.

Prior work has validated the utility of ASES to identify important themes of the physics graduate student experience that are critical for students' intentions to persist in their program through principal components analysis~\cite{Sachmpazidi22-SEP}.
More recently, the ASES dataset was used to demonstrate the NALS methodology, revealing an alternative thematic division of the survey items~\cite{Dalka22-NALS}.
The four themes found through NALS include {\it social and scholarly exploration support} ($E$), {\it mentoring and research experience} ($R$), {\it professional and academic development} ($D$), and {\it financial support} ($F$).
See a full list of the survey items in Appendix~\ref{app:nals_themes}.

In addition to the survey items, ASES data include participants' responses to a set of demographic questions, such as type of program, gender, age, number of semesters since enrollment, funding situation, established academic mentor, etc.
This additional information can be used to investigate whether the resulting themes vary between distinct respondent groups.
In this study, we chose to focus on four demographics for which the subgroups were well defined and sufficiently large to apply the NALS methodology.
In this work, we focus on four demographics: type of program (with $N_{\rm bridge}=214$ and $N_{\operatorname{nonbridge}}=167$), gender (with $N_{\rm women}=100$ and $N_{\rm men}=277$), the number of semesters since enrollment (with $N_{\rm early}=139$ for less than five semesters and $N_{\rm later}=229$ for five or more semesters), and type of support available since enrollment (with $N_{\rm research}=130$, $N_{\operatorname{nonresearch}}=108$, and $N_{\rm mixed}=143$).
We will discuss the details of each group in Sec.~\ref{ssec:net_analysis}.

%%%%%%%%%%%%%%%%%%%%%%%%%%%%%%%%%%%%%%%%%%%%%%%%%%%%%%%%%%%%%%
\subsection{Creation of the ASES backbone network}
\label{ssec:net_creation}
%%%%%%%%%%%%%%%%%%%%%%%%%%%%%%%%%%%%%%%%%%%%
In this section, we briefly describe the steps used by the NALS approach to generate networks.
NALS applies to any Likert-style survey instrument with a scale ranging from negative (disagreement) to positive (agreement) association in which sections are coded in the same direction.
The steps to create the backbone\footnote{A \textit{backbone} network is the resulting network after eliminating nonsignificant edges from the full network.} survey network, laid out in detail in Ref.~\cite{Dalka22-NALS}, can be summarized as follows:
\begin{enumerate}[topsep=2pt,itemsep=-2pt]
    \item Create a bipartite network of respondents and response selections that include all possible Likert-scale options for each question.~\footnote{A \textit{bipartite network} is one in which there are nodes of two types, with nodes of one type only linked to nodes of the opposite type.}
    \item Project the bipartite network onto response selections using the edge weights to indicate the number of respondents selecting each pair of responses.
    \item For each pair of items, build a relation matrix (a $k\times k$ matrix for $k$-level scale).
    Collapse the relation matrix to a $2 \times 2$ by first summing up response options that capture the same direction (e.g., agree and strongly agree) and then removing the neutral option (if applicable).~\footnote{The off-diagonal elements are eliminated from analysis as they indicate a lack of attitude to a given question.}
    \item Calculate the similarity score between items by first summing up the diagonal elements of the collapsed relation matrix and then subtracting the off-diagonal elements.
    A positive (negative) value indicates that the two items are answered with similar (dissimilar) responses.
    \item For a positive similarity score, calculate the direction of the association by subtracting from the sum of all mutual agrees the sum of all mutual disagrees.
    A positive (negative) association is coded as a positive (negative) temperature, indicated with a red (blue) edge.
    \item The resulting matrix defined the full survey item network.
    Use a network specification algorithm to determine the backbone survey network.
\end{enumerate}
For a more detailed explanation of NALS, along with a visualization of key steps and an example using a toy model, see Ref.~\cite{Dalka22-NALS}.

Here, we will briefly discuss the application of each step to the ASES dataset.
The 35 survey items within the ASES instrument are structured with five Likert-style response options: strongly disagree, disagree, neither agree nor disagree, agree, and strongly agree.
Given this, there are 175 unique responses available.
Since respondents can choose only one option per item, any single respondent is linked to 35 responses.
In the full dataset, this means that there will be 381 students, each connected to 35 of the total 175 responses.

The first step in building the backbone ASES network is to generate the bipartite network.
The two node types in the bipartite network in NALS are respondents and response options.
In a network built from the full ASES dataset, there are 381 student respondent nodes.
In networks built from distinct demographic groups, this number corresponds to the number of respondents in that group.
The number of unique item responses in the full dataset is 175, and it remains unchanged regardless of which demographic group is considered.
In the next step, the bipartite network is projected onto the response options.
If every possible response to each item in the dataset has been selected at least once, the resulting network has 175 nodes.
The number is lower if there are response options that were not selected by any respondent.
The survey response networks are then condensed through item relation matrices, as described in steps 2 through 4.
All resulting survey item networks have 35 nodes.
For the full dataset, the maximum resulting weight of an edge is 381 (the number of respondents).
In networks built from demographic groups, the maximum weight is lower and bounded by the number of respondents in a given group.
In addition to the similarity score, edges with a positive score are also tagged with the appropriate temperature, as described in step 5.
Finally, a locally adaptive network sparsification (LANS)~\cite{Foti11-NSN} is used on each network to determine the survey network backbone.
In this approach, significant edges for each node are identified and preserved in the network based on the empirical cumulative distribution of edge weights~\cite{Foti11-NSN}.
This ensures that all edges significant for at least one node are retained.
Depending on the distribution of edge weights in different networks, this may result in different numbers of edges in each network.

By presenting both similarity and temperature in the visualization of NALS-generated networks, we aim to show which survey items are related and in what ways.
This allows us to investigate the nature of relationships for items identified as related through similarity.
In the case of ASES, this allows us to better understand whether or not support structures are a part of graduate students' experience and how they form a cohesive set of supports in graduate programs.

The size of each demographic-based network generated by applying NALS to the ASES subset is the same.
This makes a comparison between networks fairly straightforward.
The main difference between the networks is the upper bound on the edge weights, which is determined based on the size of each demographic group.
To simplify the comparison, we normalize the weights within each network by the respective maximum weights, bringing all weights to a $(0,1)$ range.

%%%%%%%%%%%%%%%%%%%%%%%%%%%%%%%%%%%%%%%%%%%%%%%%%%%%%%%%%%%%%%
\subsection{Analysis of the network}
\label{ssec:net_analysis}
%%%%%%%%%%%%%%%%%%%%%%%%%%%%%%%%%%%%%%%%%%%%
The primary focus of this paper is to analyze the different networks that are produced when investigating demographic groups in the ASES dataset.
In this section, we describe these groups as well as the analysis tools we use to analyze the resulting networks.

In our analysis, we consider ten survey backbone networks.
As a benchmark, we use the {\it full} network created based on the complete ASES dataset.
The demographic-based networks include:
\begin{itemize}[topsep=2pt, itemsep=-2pt]
    \item two networks representing the {\it bridge} (including bridge and bridge partnership sites) and {\it nonbridge} programs,
    \item two networks representing {\it women} and {\it men} respondents (while it can be reductive~\cite{Traxler16-EGP}, we use the gender binary of women and men to separate respondents as the number of nonbinary and nonreporting respondents was too low to be analyzed in this work),
    \item two networks created based on the number of semesters since enrollment reported by respondents, with {\it early} indicating four or fewer semesters (typically when students are completing coursework) and {\it later} indicating five or more semesters (typically when students are focused on research), and  
    \item three networks created based on the type of funding students have been supported by during their time in graduate school, with {\it research} indicating only fellowships and/or research assistantships (RA), {\it nonresearch} indicating teaching assistantships (TA) and/or loans, and {\it mixed} indicating a combination of fellowships, RAs, TAs, and/or loans.
\end{itemize}

To describe networks globally, we look at the total number of nodes, edges, and components.
Identifying the number of components in a network, where a \textit{component} is defined as a subset of nodes such that there is at least one path between any two members in the set and no paths exist to the rest of a network, can help explain the network connectivity.
In addition to the features of a network we can directly count, density characterizes how connected a network is.
A network's density describes how many edges exist out of the total possible number of edges that could have existed.

To compare networks, we first focus on the structural similarity based on nodes and edges.
We use the {\it node degree cosine} (NDC) similarity to measure the extent to which the degree values of nodes are the same and the {\it edge existence Jaccard} (EEJ) similarity for the edges~\cite{Brodka18-QLS}.
A node's degree is the count of edges connected to that node.
Both measures range from 0 to 1, with 1 indicating perfect matching in terms of nodes' degrees for NDC and existing edges for EEJ.

NDC is calculated by taking the list of nodes' degrees for two networks ($A$ and $B$) and calculating the cosine similarity between them:
\begin{equation}
    \text{NDC} = \frac{\sum\nolimits_{i=1}^N \mathcal{C}_{D}^{A}(i) \mathcal{C}_{D}^{B}(i)}{\sqrt{\sum\nolimits_{i=1}^N \mathcal{C}_{D}^{A}(i)^2}\sqrt{\sum\nolimits_{i=1}^N \mathcal{C}_{D}^{B}(i)^2}},
\end{equation}
where $\mathcal{C}_{D}^{A(B)}(i)$ represents the degree of node $i$ within network $A$ ($B$) [see Eq.~(\ref{eq:degree})]~\cite{Leydesdorff05-SMA}.
EEJ is calculated by dividing the size of the intersection of two edge lists (i.e., the lists of all existing edges) by the size of the union of those edge lists and is given by
\begin{equation}
    \text{EEJ} = \frac{|E_{A} \cap E_{B}|}{|E_{A} \cup E_{B}|},
\end{equation}
where $E_{A(B)}$ is the edge list of network $A$ ($B$)~\cite{Ivchenko98JST}.

When interpreting, these measures of similarity must be understood in relation to each other.
We did not find agreed-upon cutoffs for either EEJ or NDC, as different types of networks with varying edge density and structure will have different expectations for each of them.
Thus, when we discuss a particular EEJ or NDC measure as being ``low'' or ``high,'' we do so in reference to the other comparisons made within this study.

The main focus of our study is on the {\it partitioning} of the networks.
Network partitioning is how the nodes of the network are separated into {\it clusters}~\cite{Newman06-MCS}.
A cluster is a group of nodes that share stronger connections to other nodes within that group than outside of the group.
In our previous work, we used cluster analysis to identify themes within the ASES instrument~\cite{Dalka22-NALS}.
The {\it modularity} of partitioning quantifies how well the network is separated into smaller clusters.

In this study, we use the weighted undirected definition of modularity defined as
\begin{equation}
Q = \frac{1}{2m}\sum\nolimits_{i,j} \bigg[w_{ij} - \frac{\mathcal{C}_{S}(i)\, \mathcal{C}_{S}(j)}{2m}\bigg]\,\delta(C_i,C_j),
\end{equation}
where $w_{ij}$ represents the weight of a tie between nodes $i$ and $j$, $\mathcal{C}_{S}(i)=\sum_k w_{ik}$ $(\mathcal{C}_{S}(j)=\sum_k w_{jk})$ summed over all nodes directly connected to $i$ ($j$) represents the strength of node $i$ ($j$), $C_{i (j)}$ indicates the community to which node $i$ ($j$) belongs, and $m = \frac{1}{2}\sum_{i,j}w_{ij}$~\citep{Newman04-AWN}.
The delta function, $\delta(C_i,C_j)$, equals $1$ when $C_i = C_j$ and $0$ otherwise.
The modularity ranges from $-1$ to $1$ and compares the relative density of ties within communities and between communities.
A positive value indicates a partitioning in which the ties within communities are more prevalent than those between communities.
To create each network partitioning, we use the hierarchical clustering algorithm, commonly referred to as the fast-greedy algorithm~\cite{Clauset04-FCS}.
This algorithm optimizes for modularity and creates a hierarchical ordering that partitions the network into clusters.

To compare the partitioning of networks, we use the {\it purity} measure~\cite{Ghawi22-CMS}.
Purity is a common metric for comparing two partitions by comparing how well two partitionings overlap cluster by cluster.
Cluster purity quantifies the extent to which a cluster $\mathcal{S}_m^A$ from partitioning A contains nodes from only a single cluster $\mathcal{S}_n^B$ from partitioning B.
The overall purity comparing two partitionings is defined as the sum of all individual cluster purities weighted by the cluster size~\cite{Zhao01CFC}.
Purity ranges from 0, indicating no overlap in partitions, to 1, indicating a perfect matching.
While traditional cluster comparison is asymmetric, treating one partitioning as the base and the other as the variant, we use a two-way comparison proposed by Ghawi and Pfeffer~\cite{Ghawi22-CMS}.
The two-way purity is defined as a harmonic mean between the two purities calculated by swapping which partitioning is considered the base.
Similar to interpreting EEJ and NDC measures, we compare ``low'' and ``high'' purity within the context of the demographic-based and full networks rather than comparing to other networks that have different types of partition differences.

Finally, to compare the nodes that bridge between clusters, we use two centrality measures: the total degree $\mathcal{C}_D$ and betweenness $\mathcal{C}_B$.
The total degree quantifying the number of immediate connections to a given node is a local, node-level measure of connectivity.
It is defined as
\begin{equation}\label{eq:degree}
    \mathcal{C}_D(i) = \sum\nolimits_{j=1}^N x_{ij},
\end{equation}
where $N$ is the network size and $x_{ij}$ is $1$ when there is an edge between note $i$ and $j$ and $0$ otherwise.
The betweenness quantifying the number of times a node acts as a ``bridge'' along the shortest path linking two other nodes is a global measure of connectivity. 
It is defined as 
\begin{equation}
    \mathcal{C}_B(i) = \sum\nolimits_{m\neq i \neq n} \frac{\ell_{mn}^{(i)}}{\ell_{mn}},
\end{equation}
where $\ell_{mn}^{(i)}$ is the number of shortest paths linking nodes $m$ and $n$ that pass through node $i$ and $\ell_{mn}$ is the total number of shortest paths linking nodes $m$ and $n$.
While the total degree captures the size of the network of immediate connections, the betweenness captures the importance of a node's position within a whole network. 
In other words, degree quantifies how well connected a given node is and betweenness identifies nodes that connect clusters that would otherwise split into subcomponents.

%%%%%%%%%%%%%%%%%%%%%%%%%%%%%%%%%%%%%%%%%%%%%%%%%%%%%%%%%%%%%%
\subsection{Statistical analysis and visualization}
\label{ssec:stat-viz}
%%%%%%%%%%%%%%%%%%%%%%%%%%%%%%%%%%%%%%%%%%%%
To test the robustness of the network partitioning against small perturbations, we employ statistical bootstrapping techniques~\cite{Efron94-IB}.
Given how our networks are created, we choose to bootstrap at the ASES level survey data rather than resampling directly from the networks~\cite{Rosvall10-MCN}.
For each performed test, the bootstrapping consists of three steps.
First, an appropriate subset of the ASES data is selected based on the characteristic of interest (e.g., gender or program type). 
Then, a hypothetical dataset is drawn at random from the actual students' responses included in the subset.
A random drawing with replacement is employed to ensure that the hypothetical datasets are the same size as the subset.
This means that any given response from the original subset may be selected more than once or might be omitted from a given hypothetical dataset. 
Each hypothetical dataset is then used to create a new backbone network for which the partitioning is determined in the third step.

The bootstrapping process is repeated $1,000$ times for the full backbone network and for each demographic group to ensure saturation (see Appendix~\ref{app:convergence} for bootstrapping convergence analysis).
Once bootstrapping is completed, we check how frequently nodes are assigned to the cluster determined from the complete dataset for a given group.
We choose as reference the $50~\%$ threshold to indicate which nodes are more often than not grouped into their original cluster.
We do not use this threshold as an absolute cutoff to determine cluster stability but rather as an indicator of the emerging thematic groupings based on a high frequency of clustering. 
Nodes below this threshold are either closely related to a different cluster or are marginally related to multiple clusters.

Bootstrapped networks are also used to test the effect size of each measure by comparing the demographic-based networks to the full network.
The effect size is relevant to understanding the differences represented in Table~\ref{tab:mod_meas}.
We calculate the effect size using Cohen's $d$ measure~\cite{Cohen13} defined as
\begin{equation}\label{eq:cohen_d}
    d = \frac{|M_{\gamma({\rm A,B})} - M_{\gamma({\rm A,C})}|}{\sigma_p},
\end{equation}
where $M_{\gamma({\rm A,B})}$ and $M_{\gamma({\rm A,C})}$ are the means of a network measure $\gamma$ comparing networks A and B and networks A and C, respectively, and $\sigma_p$ is the pooled standard deviation (the weighted average of the two standard deviations).
Comparisons are always made between relevant demographic-based networks and the full network.
Cohen's $d$ quantifies how strong a difference between two measures is and is commonly divided into small ($0.2 \leq d < 0.5$), medium ($0.5 \leq d < 0.8$), and large ($0.8 \leq d$)~\cite{Cohen13}.
The Cohen's $d$ values for all comparisons considered in this work are presented in Table~\ref{tab:cohens_d} in Appendix~\ref{app:effect_size}.

To investigate the dependence of network comparison measures (NDC, EEJ, and purity) on the size of the sampled population, we ran $2,000$ bootstrapped tests for sample sizes in increments of 50, up to the full size of the dataset.
We find that while there is some relationship between the size of the population sampled and these three network comparison measures, it does not fully explain the differences we see between demographic-based networks and the full network.
See Appendix~\ref{app:sample-size} for more details.

All analyses presented in this work are carried out using the {\it igraph} package in {\it R} \cite{igraph, R}.
Employing the LANS algorithm, we use a level of significance $\alpha = 0.05$. 
The network visualization is created using the open-source software \textsc{cytoscape} \cite{cytoscape}.

%%%%%%%%%%%%%%%%%%%%%%%%%%%%%%%%%%%%%%%%%%%%%%%%%%%%%%%%%%%%%%
\section{Results}\label{sec:results}
\begin{figure}[b]
    \centering
    \includegraphics{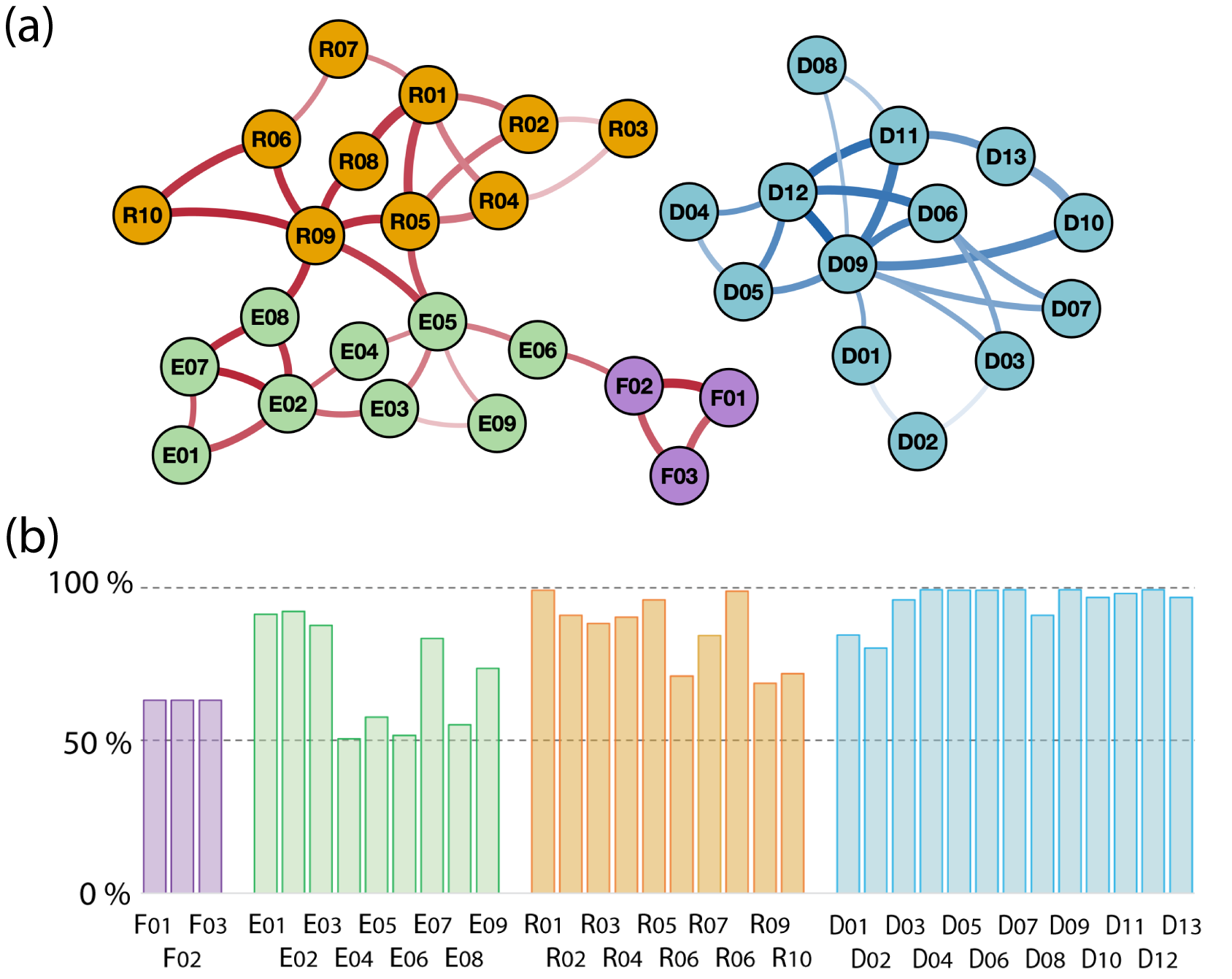}
    \caption{(a) The network created based on the full ASES dataset. 
    The four themes are highlighted with different colors. 
    Positive temperature edges are indicated in red and negative temperature edges are indicated in blue.
    (b) The bootstrapping frequency plot for the full ASES dataset, grouped by network clustering.}
    \label{fig:networks-full}
\end{figure}

%%%%%%%%%%%%%%%%%%%%%%%%%%%%%%%%%%%%%%%%%%%%%%%%%%%%%%%%%%%%%%
In our previous work, we found four clusters within the network built from respondents to the ASES instrument using NALS.
The clusters that emerged from the full ASES dataset, shown in Fig.~\ref{fig:networks-full}(a), were named {\it social and scholarly exploration support} ($E$, shown in green), {\it mentoring and research experience} ($R$, shown in orange), {\it professional and academic development} ($D$, shown in blue), and {\it financial support} ($F$, shown in purple).

%%%%%%%%%%%%%%%%%%%%%%%%%%%%%%%%%%%%%%%%%%%%
%%%%%%%%%%%%%%%%%%%%%%%%%%%%%%%%%%%%%%%%%%%%

In this work, we focus on two aspects of the NALS analysis of ASES data: the partitioning variability between different demographic groups split along a given dimension into mutually exclusive subgroups and the stability of the emerging clusters.
The leading questions of the current study are as follows:
\begin{enumerate}[topsep=2pt, itemsep=-5pt] %nosep
    \item How stable is the NALS clustering?
    \item How, if at all, do the clusters change when calculated for well-defined respondent groups?
    \item What do the differences between subgroup clustering mean for student experiences?
\end{enumerate}

For the full ASES network, we find that all clusters are stable at the $50~\%$ level, as shown in Fig.~\ref{fig:networks-full}(b). 
The least stable cluster is $E$, which has the most edges extending to other clusters in the network, while the most stable cluster is $D$.

In the following section, we focus on the four demographic splits discussed in Sec.~\ref{ssec:net_analysis}.
In our analysis, we compare the ASES network splits with the full ASES network and with each other.
First, we make comparisons based on the NDC and EEJ similarities.
Next, we compare networks based on the clustering structure of each network, quantified by the purity.
Finally, to investigate the stability of the partitions, we use the bootstrapping method, as described in Sec.~\ref{ssec:stat-viz}.
For each test, we compare the frequency of nodes' assignment to clusters determined based on the relevant dataset. 
We also use the NDC and EEJ similarity measures to compare the bootstrapped networks with the reference ones.
\vspace{-10pt}

%%%%%%%%%%%%%%%%%%%%%%%%%%%%%%%%%%%%%%%%%%%%%%%%%%%%%%%%%%%%%%
\subsection{Network measures and comparisons}
%%%%%%%%%%%%%%%%%%%%%%%%%%%%%%%%%%%%%%%%%%%%%%%%%%%%%%%%%%%%%%
We begin by presenting the basic network descriptors for all networks considered in this work, as seen in Table~\ref{tab:net_meas}.
Since the sparsification algorithm affects only the number of edges ($N_E$) but not the number of nodes ($N_N)$, all networks have 35 nodes.
The number of edges varies as the distribution of edge weights for nodes in each network may differ.
The number of edges is on average $M_{N_E}=54.9(2.6)$,~\footnote{We use a notation value(uncertainty) to express uncertainties, for example, $1.5(6)~\si{\centi\meter}$. 
All uncertainties herein reflect the uncorrelated combination of single-standard deviation statistical and systematic uncertainties.} ranging from $50$ edges in the women's network to $61$ edges in the bridge network.

The full, nonbridge, men, early, and later backbone networks have two components while the remaining networks have a single large component.
In all two-component networks, the larger component [$M_{S(c_1)}=21.8(4)$] consists of the F, E, and R themes connected via positive temperature edges.
The nodes in the smaller [$M_{S(c_1)}=13.2(4)$], predominantly D-themed component, are connected via negative temperature edges.
Finally, the average component density is $M_{\Delta(c_1)}=0.15(1)$ and $M_{\Delta(c_2)}=0.26(2)$ for the two-component networks and $M_{\Delta(c_1)}=0.09(1)$ for the single-component networks.
Due to the higher density, the smaller component is often grouped as a singular cluster.

Within each network, different nodes become central to the structure (see Table~\ref{tab:cent_meas} in Appendix~\ref{app:centralities} for a comparison of centrality measures for all networks considered in this work).
In the full network, the node with the largest degree is $D_{09}$ [$\mathcal{C}_D(D_{09}) = 9$] while nodes $E_{05}$ and $R_{09}$ have the highest betweenness measures [$\mathcal{C}_B(E_{05}) = 106$ and $\mathcal{C}_B(R_{09}) = 75$, respectively].
In the partitioning of the bootstrapped full network, $D_{09}$ is almost always placed into the $D$ cluster while $E_{05}$ and $R_{09}$ are among the nodes most likely to move clusters.
This suggests that the betweenness metric can help us identify nodes that may be unstable in partitions of bootstrapped networks.

Table~\ref{tab:mod_meas} shows both NDC and EEJ for all pairwise comparisons.
When comparing networks built from demographic splits to the full network, the NDC ranges from 0.81 (research to full) to 0.98 (nonbridge to full, men to full, and later to full), with an average of $M_{\text{NDC}} = 0.94(5)$.
The NDC between demographic-based networks tends to be somewhat lower, with $M_{\text{NDC}} = 0.87(7)$.
The biggest difference is observed for the funding support-related splits, with $\text{NDC} = 0.76$ for research vs nonresearch comparison and $\text{NDC} = 0.79$ for research vs mixed.
The relatively high NDC values are consistent with the sparsification process, as each node will only have a few ties to other nodes.
Cohen's $d$ indicates no large or medium effect sizes for the comparisons to the full network.
There are small effects seen in the NDC for each funding support-related split compared with the full network.

%%%%%%%%%%%%%%%%%%%%%%%%%%%%%%%%%%%%%%%%%%%%
\begin{table}[t]
\renewcommand{\arraystretch}{1.02}
\renewcommand{\tabcolsep}{2pt}
\caption{Basic network descriptors for the full network and all subnetworks: 
number of edges ($N_E$), 
the number of components ($N_c$), and the size [$S(c_i)$ for $i=1,2$], and density [$\Delta(c_i)$, $i=1,2$] of the components.
\label{tab:net_meas}%
}
\begin{ruledtabular}
\begin{tabular}{lcccccc}
 & $N_E$ & $N_c$ & $S(c_1)$ & $S(c_2)$ & $\Delta(c_1)$ & $\Delta(c_2)$ \\ \hline
Full         & 55 & 2 & 22 & 13 & 0.15 & 0.27 \\ \hline
Bridge       & 61 & 1 & 35 & $\cdots$ & 0.10 & $\cdots$ \\
Nonbridge   & 54 & 2 & 21 & 14 & 0.15 & 0.23 \\ \hline
Women        & 50 & 1 & 35 & $\cdots$ & 0.08 & $\cdots$ \\
Men          & 55 & 2 & 22 & 13 & 0.14 & 0.28 \\ \hline
Early        & 56 & 2 & 22 & 13 & 0.16 & 0.26 \\
Later        & 56 & 2 & 22 & 13 & 0.15 & 0.27 \\ \hline
Research     & 54 & 1 & 35 & $\cdots$ & 0.09 & $\cdots$ \\
Non-research & 53 & 1 & 35 & $\cdots$ & 0.09 & $\cdots$ \\
Mixed        & 55 & 1 & 35 & $\cdots$ & 0.09 & $\cdots$ \\
\end{tabular}
\end{ruledtabular}
\end{table}
%%%%%%%%%%%%%%%%%%%%%%%%%%%%%%%%%%%%%%%%%%%%

The EEJ ranges from 0.28 (women to full) to 0.75 (men to full), with an average of $M_{\text{EEJ}} = 0.5(2)$ when comparing networks built from demographic splits to the full network.
Similar to the NDC, the EEJ between networks built from demographic splits tends to be lower, with $M_{\text{EEJ}} = 0.30(7)$.
The low EEJ indicates networks that are unique when compared to each other.
The most dissimilar networks are research and nonresearch, with $\text{EEJ} = 0.19$, indicating that these demographics result in almost entirely different connectivity.
The differences we see in Table~\ref{tab:mod_meas} are reflected by Cohen's $d$, all of which have a large effect size when comparing the EEJ of demographic-based networks to the full network.
This means that each network within the demographic-based splits contains a unique set of edges, indicating that the experiences of support structures are connected in different ways.

The purity of partitionings between demographic splits and the full network ranges from 0.74 (women to full and nonresearch to full) to 0.87 (early to full), with an average of $M_{\text{purity}} = 0.80(4)$.
The purity between networks built from demographic splits is slightly lower, with $M_{\text{purity}} = 0.73(6)$ and a minimum of 0.61 for research vs nonresearch comparison.
Purities for all comparisons considered in this work are presented in Table~\ref{tab:mod_meas}.
The effect sizes of these differences in purity are also quantified using Cohen's $d$.
The differences in the women network to full network and the men network to full network have a large effect size.
Similarly, there is a large effect size in the differences of purity between the early network to full network and the later network to full network.
The purity of demographic-based networks to the full networks has a large effect size also when comparing men and women networks as well as early and later networks.
This means that the partition of the women network is measurably different than that of men and similar for the early semester respondents to the later semester respondents.
For the purity between the research network to full network and the nonresearch network to full network, and for the purity between the research network to full network and the mixed funding network to full network, the effect size is medium.
The medium effect size of the purity in the funding-based network indicates a similar, but less variable, difference between these network partitions.

%%%%%%%%%%%%%%%%%%%%%%%%%%%%%%%%%%%%%%%%%%%%
\begin{table}[t]
\renewcommand{\arraystretch}{1.02}
\renewcommand{\tabcolsep}{2pt}
\caption{\label{tab:mod_meas}
A summary of the structural similarity measures between networks A and B in terms of nodes' degree (NDC), edges present in networks (EEJ), and partitioning of networks (purity).}
\begin{ruledtabular}
\begin{tabular}{llccc}%c}
Network A & Network B  & NDC & EEJ & Purity \\ \hline
Bridge       & Full         & 0.95 & 0.55 & 0.83 \\ 
Nonbridge   & Full         & 0.98 & 0.58 & 0.80 \\ 
Bridge       & Nonbridge   & 0.93 & 0.37 & 0.73 \\ \hline
Women        & Full         & 0.96 & 0.28 & 0.74 \\
Men          & Full         & 0.98 & 0.75 & 0.84 \\ 
Women        & Men          & 0.96 & 0.40 & 0.75 \\ \hline
Early        & Full         & 0.91 & 0.39 & 0.87 \\ 
Later        & Full         & 0.98 & 0.73 & 0.82 \\ 
Early        & Later        & 0.89 & 0.29 & 0.77 \\ \hline
Research     & Full         & 0.81 & 0.42 & 0.79 \\ 
Nonresearch & Full         & 0.91 & 0.33 & 0.74 \\ 
Mixed        & Full         & 0.96 & 0.59 & 0.80 \\ 
Research     & Nonresearch & 0.76 & 0.19 & 0.61 \\
Research     & Mixed        & 0.79 & 0.30 & 0.76 \\ 
Nonresearch & Mixed        & 0.86 & 0.24 & 0.76 \\  
\end{tabular}
\end{ruledtabular}
\end{table}
%%%%%%%%%%%%%%%%%%%%%%%%%%%%%%%%%%%%%%%%%%%%

%%%%%%%%%%%%%%%%%%%%%%%%%%%%%%%%%%%%%%%%%%%%
\begin{figure*}[ht!]
    \centering
    \includegraphics[width=0.96\textwidth]{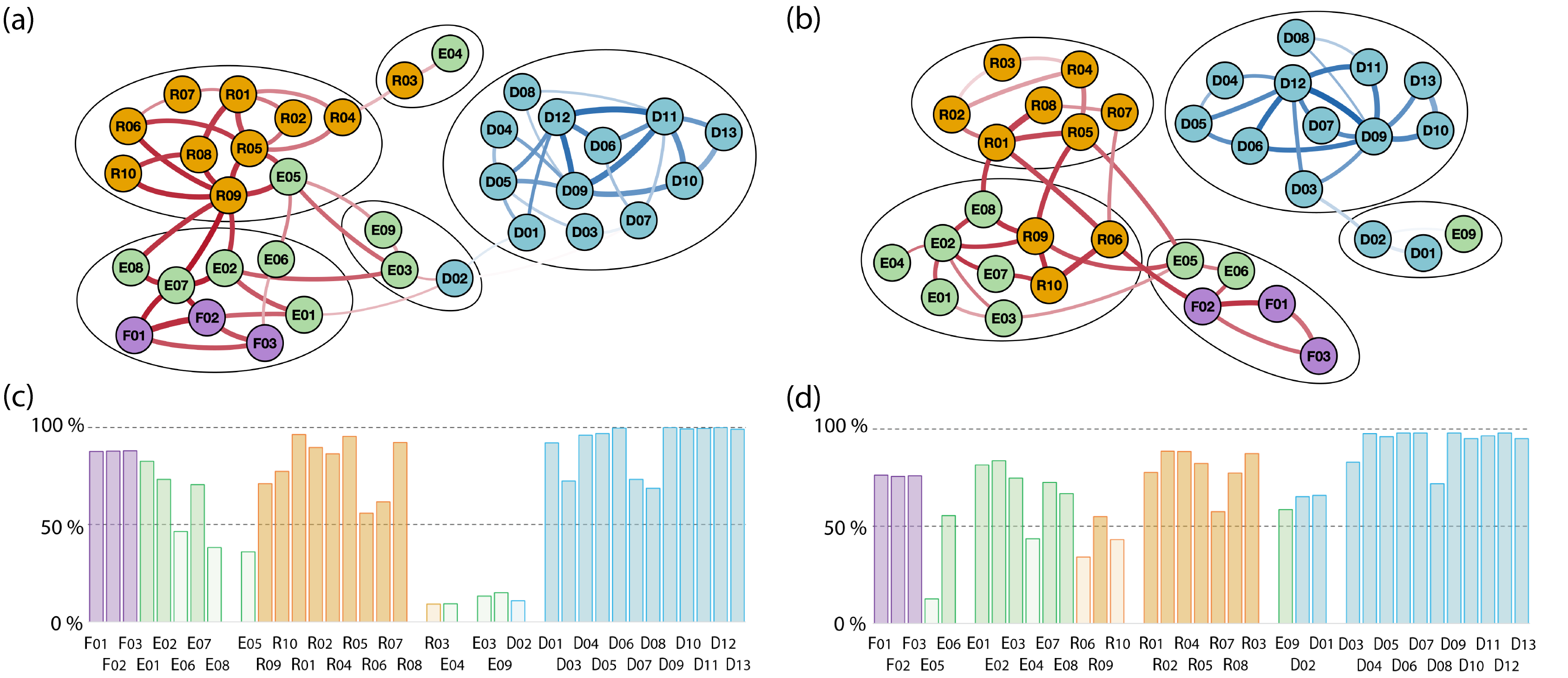}
    \caption{Plots and histograms for investigation by program type: the networks created based on the respondents from (a) bridge and (b) nonbridge programs; the frequency plots for the (c) bridge and (d) nonbridge networks bootstrapping.
    In the network plots, the four original themes are highlighted with different colors. 
    Positive temperature edges are indicated in red and negative temperature edges are indicated in blue. 
    New clusters are indicated by the circled regions. 
    Each histogram is grouped by the clusters in the respective network, with nodes not reaching 50~\% stability made transparent.
    }
    \label{fig:networks-programs}
\end{figure*}
%%%%%%%%%%%%%%%%%%%%%%%%%%%%%%%%%%%%%%%%%%%%
\vspace{-10pt}

%%%%%%%%%%%%%%%%%%%%%%%%%%%%%%%%%%%%%%%%%%%%%%%%%%%%%%%%%%%%%%
\subsection{Exploring experiences in the bridge and nonbridge programs}
%%%%%%%%%%%%%%%%%%%%%%%%%%%%%%%%%%%%%%%%%%%%
The first dimension that we split the ASES dataset on is the type of program the respondents are enrolled in.
The network that results from the respondents in bridge programs (henceforth known as the {\it bridge network}) is partitioned into five clusters, as seen in Fig.~\ref{fig:networks-programs}(a).
The network that results from the respondents in nonbridge programs (the {\it nonbridge network}) is partitioned into five different clusters, as seen in Fig.~\ref{fig:networks-programs}(b).

The NDC similarity is very high for all three network comparisons we consider, as seen in Table~\ref{tab:mod_meas}. 
This is expected as through the sparsification process, each node will only have a few edges that are identified as important, resulting in a reasonably consistent degree distribution.
Where differences become clear is when comparing which edges exist in the networks.
When the full network is compared to both bridge and nonbridge programs, we see that just over half of the edges are shared, with $\mathrm{EEJ}=0.55$ for bridge vs full network and $\mathrm{EEJ}=0.58$ for nonbridge vs full network.
When comparing the bridge program network and the nonbridge program network, we find that only about one-third of the edges are shared.
This indicates that the networks built by respondents in bridge programs are unique compared to those built by respondents in nonbridge programs.

%%%%%%%%%%%%%%%%%%%%%%%%%%%%%%%%%%%%%%%%%%%%
\begin{table}[b]
\renewcommand{\arraystretch}{1.02}
\renewcommand{\tabcolsep}{2pt}
\caption{\label{tab:mod_meas_boot}
A summary of the structural similarity measures (NDC and EEJ), the network partitioning (purity), and the number of clusters ($N_{\mathcal{S}}$) between the original demographic-based and the bootstrapped networks.
The averages are calculated based on $N=1,000$ iterations of bootstrapping.}
\begin{ruledtabular}
\begin{tabular}{lcccc}
Network &  NDC & EEJ & Purity & $N_{\mathcal{S}}$ \\ 
\hline
Bridge       & 0.95(2) & 0.50(6) & 0.77(6) & 4.4(7)\\ 
Nonbridge   & 0.95(2) & 0.55(6) & 0.79(7) & 4.9(7)\\ \hline
Women        & 0.94(2) & 0.47(6) & 0.72(6) & 5.1(8)\\ 
Men          & 0.95(2) & 0.59(6) & 0.81(6) & 4.2(8) \\ \hline
Early        & 0.92(2) & 0.45(6) & 0.76(8) & 4.7(9) \\ 
Later        & 0.95(2) & 0.53(6) & 0.82(6) & 4.3(7) \\ \hline
Research     & 0.92(3) & 0.47(6) & 0.71(7) & 4.5(8) \\ 
Nonresearch & 0.93(2) & 0.43(6) & 0.73(6) & 5.0(8) \\ 
Mixed        & 0.95(2) & 0.52(6) & 0.76(6) & 4.6(7) \\  
\end{tabular}
\end{ruledtabular}
\end{table}
%%%%%%%%%%%%%%%%%%%%%%%%%%%%%%%%%%%%%%%%%%%%

When taking into account the purity comparison metric, we see that both the bridge and nonbridge program networks have quite similar partitions as the full network.
The purity between the program-based network partitions is somewhat lower, indicating that there are some unique differences that the bridge and nonbridge program networks capture.

%%%%%%%%%%%%%%%%%%%%%%%%%%%%%%%%%%%%%%%%%%%%
\begin{figure*}[!tb]
    \centering
    \includegraphics[width=0.96\textwidth]{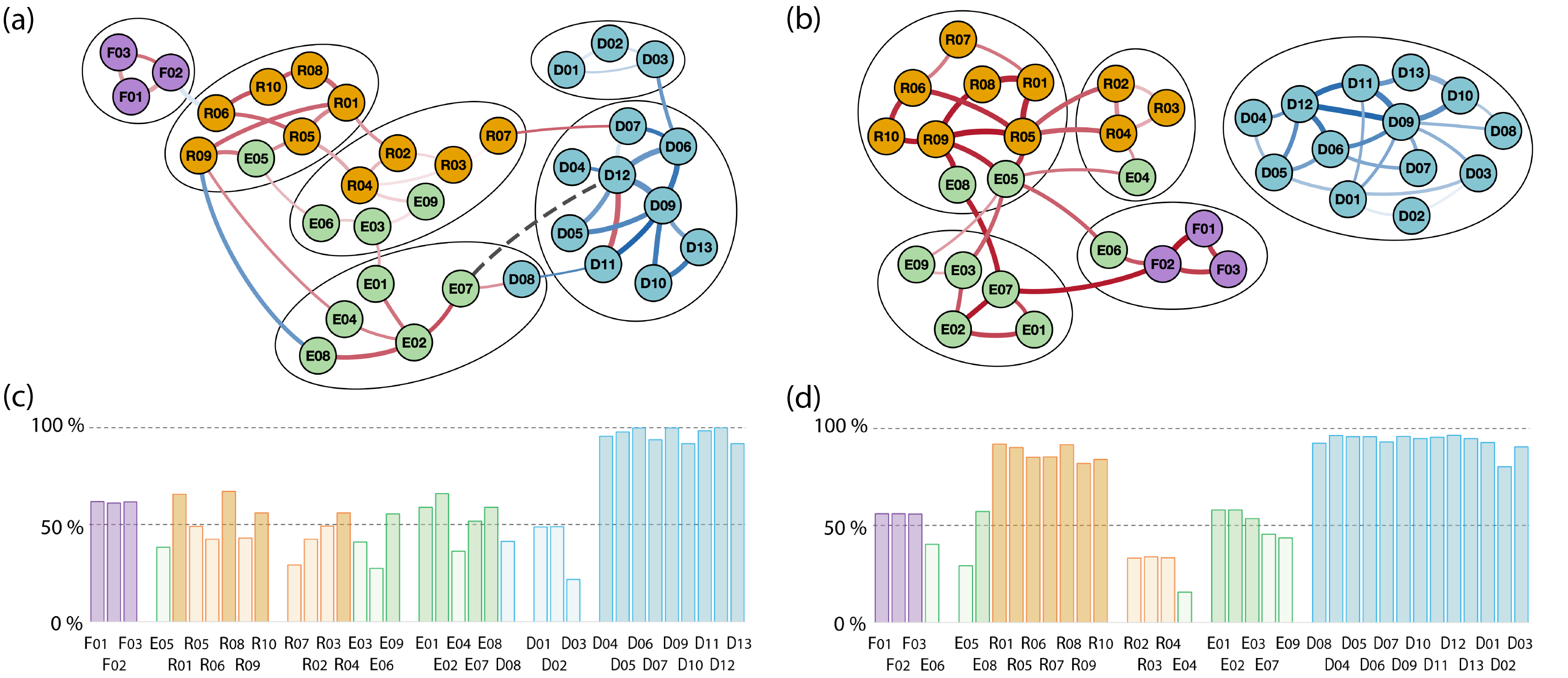}
    \caption{Plots and histograms for investigation by respondent gender: the networks created based on the (a) women and (b) men respondents; the frequency plots for the (c) women and (d) men networks bootstrapping. 
    In the network plots, the four original themes are highlighted with different colors. 
    Positive temperature edges are indicated in red and negative temperature edges are indicated in blue. 
    New clusters are indicated by the circled regions. 
    Each histogram is grouped by the clusters in the respective network, with nodes not reaching 50~\% stability made transparent.}
    \label{fig:networks-gender}
\end{figure*}
%%%%%%%%%%%%%%%%%%%%%%%%%%%%%%%%%%%%%%%%%%%%

To determine the partitions' stability, we consider the structural similarity measures (NDC and EEJ) and the partitioning purity between the reference demographic-based networks and the bootstrapped networks, see Table~\ref{tab:mod_meas_boot}.
The degree distribution for both bridge and nonbridge networks is consistently high between the bootstrapping iterations, with 
$\mathrm{NDC}=0.95(2)$ for both networks.
However, only about half of the edges that make up this network remain the same between samplings, with $\mathrm{EEJ}=0.50(6)$ and $\mathrm{EEJ}=0.55(6)$ for the bridge and nonbridge program, respectively.
The edge variability results in a slightly less consistent clustering, as confirmed by the purity measure.

The partitioning of the bridge program network is depicted in Fig.~\ref{fig:networks-programs}(a).
The three bigger clusters present in the bridge program network---$D$, $R$-like, and $E+F$---are quite persistent in the bootstrapped networks (except for two nodes in the $E+F$ cluster).  
The two smaller clusters turn out to be very infrequent in the bootstrapped networks, as seen in Fig.~\ref{fig:networks-programs}(c).
The $D_{02}$ node of the $D_{02}$, $E_{03}$, and $E_{09}$ cluster is frequently absorbed by the large $D$ cluster. 
The two other nodes, $E_{03}$ and $E_{09}$, are most frequently brought into a cluster with $E_{05}$ and $E_{06}$ and these four nodes are either in their own clusters or tend to bounce between the majority $R$ cluster and the $E$+$F$ cluster.
The other small cluster, $R_{03}$ and $E_{04}$, is absorbed by the large $R$-like cluster.
Finally, the $E_{08}$ is usually moved into a cluster with $R_{09}$ and $R_{10}$, which is the $R$-like cluster.

For the nonbridge network, shown in  Fig.~\ref{fig:networks-programs}(b), the $D$ and $R$-like clusters are most consistent across the samples.
The cluster that groups nodes from $R$ and $E$ has a few nodes that are likely to be reassigned to different clusters.
The $R_{06}$ and $R_{10}$ both tend to be more often clustered with the other $R$ nodes while the $E_{04}$ tends to move between clusters during bootstrapping.
The $E_{05}$ node (originally assigned to the $E+F$ cluster) has strong connections not just within that cluster but also bridges into two other clusters and tends to get absorbed by one of them.
Unlike the bridge program network, where the small clusters often disappear during sampling, neither of the smaller clusters gets absorbed by the larger ones in the nonbridge network.
Even the smallest $D_{01}$, $D_{02}$, and $E_{09}$ cluster persisted under sampling.
Interestingly, $D_{02}$ folds back into the $D$ cluster in the bridge network.
We can see how the low similarity score for this node, along with the varied temperatures of the edges connected to it, in the bridge network inform how this particular survey item is clustered.
\vspace{-10pt}

%%%%%%%%%%%%%%%%%%%%%%%%%%%%%%%%%%%%%%%%%%%%%%%%%%%%%%%%%%%%%%
\subsection{Exploring experiences between women and men respondents}
%%%%%%%%%%%%%%%%%%%%%%%%%%%%%%%%%%%%%%%%%%%%
The second dimension we consider in this work is gender.
The network that results from women respondents (henceforth known as the {\it women network}) is partitioned into six clusters, as seen in Fig.~\ref{fig:networks-gender}(a).
The network that results from men respondents (the {\it men network}) is partitioned into five different clusters, as seen in Fig.~\ref{fig:networks-gender}(b).

%%%%%%%%%%%%%%%%%%%%%%%%%%%%%%%%%%%%%%%%%%%%
\begin{figure*}[ht!]
    \centering
    \includegraphics[width=0.96\textwidth]{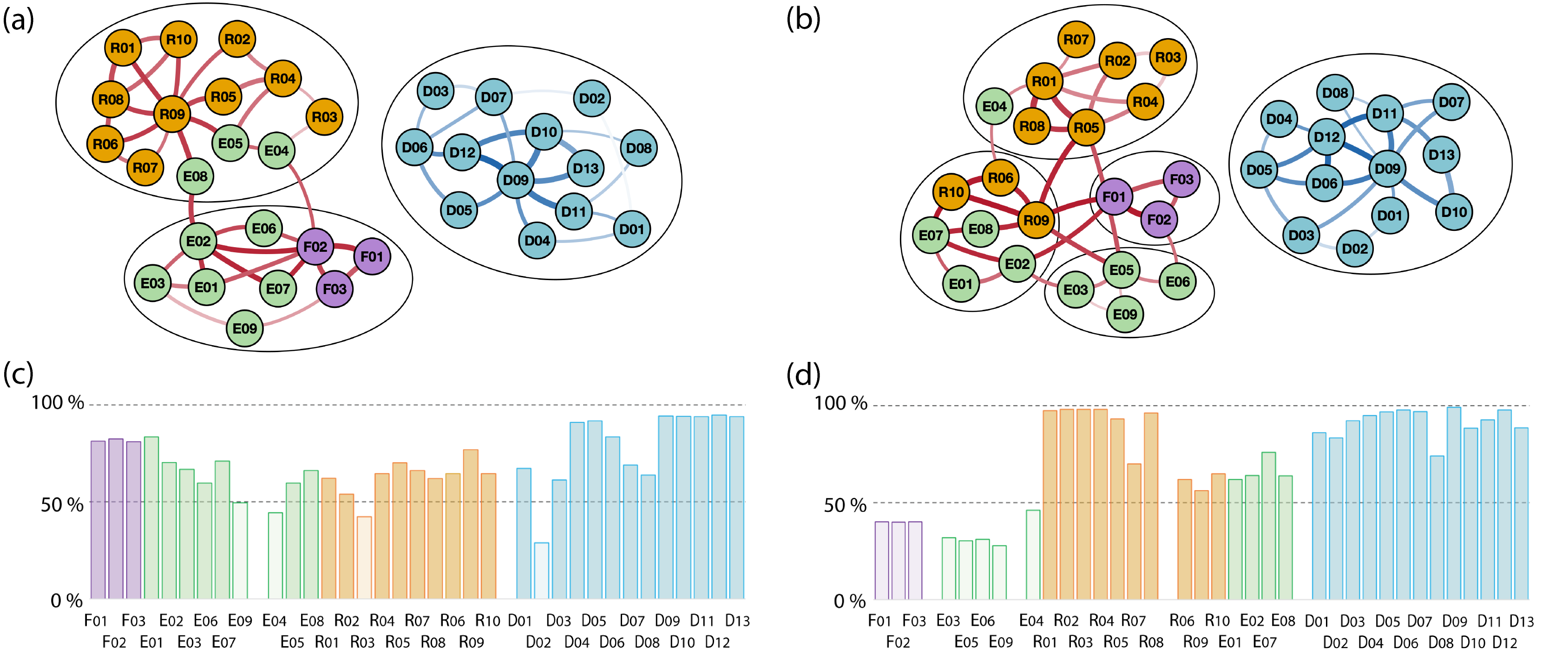}
    \caption{Plots and histograms for investigation by respondent's indicated semester: the networks created based on the respondents (a) early and (b) later in their graduate studies; the frequency plots for the (c) early and (d) later networks bootstrapping.
    In the network plots, the four original themes are highlighted with different colors.
    Positive temperature edges are indicated in red and negative temperature edges are indicated in blue.
    New clusters are indicated by the circled regions.
    Each histogram is grouped by the clusters in the respective network, with nodes not reaching 50~\% stability made transparent.
    }
    \label{fig:networks-semester}
\end{figure*}
%%%%%%%%%%%%%%%%%%%%%%%%%%%%%%%%%%%%%%%%%%%%

Similar to the program-based split, the NDC similarity is very high for the gender networks, as seen in Table~\ref{tab:mod_meas}.
However, the three pairwise network comparisons result in very different EEJ similarities.
The comparison between the full network and the network created by women respondents has a very low $\mathrm{EEJ}=0.28$.
When comparing the network created by men respondents to the full network, we see that there are many more shared edges, with the highest $\mathrm{EEJ}=0.75$.
In fact, the men network is the most similar to the full network out of all subgroups considered in this work while the women network is the least similar.
This indicates that the network structures---and thus the resulting partitioning of the networks---vary significantly for the two gender-based groups of respondents.
This is further confirmed by the purity measure.
While the difference in similarities with the full network may be partially due to the difference in sample sizes, it does not fully explain the large difference we see in these comparisons (see Appendix~\ref{app:sample-size} for more details).

The partitioning of the women network results in four fairly equally sized clusters consisting of seven to nine nodes and two smaller ones, each consisting of three nodes, see Fig.~\ref{fig:networks-gender}(a).
However, only the largest $D$ cluster and the $F$ cluster are persistent in the bootstrapped networks, as seen in Fig.~\ref{fig:networks-gender}(c). 
All of the other clusters identified change depending on which subset of women respondents we sample from.
Additionally, while the $D$ cluster is stable for the women network, the structure of internal edges suggests further investigation might be necessary to better understand how the support structures included in cluster $D$ are experienced by women.
Specifically, $D_{12}$ creates unique connections within the $D$ cluster as we can identify it as a polarizing survey item for the women network.
It has both positive and negative temperature similarities, as well as a strong dissimilar connection.
In the men network, this particular survey item behaves more consistently with other $D$ nodes.

The partitioning of the men network is very similar to that of the full network, though there are five rather than four clusters, see Fig.~\ref{fig:networks-gender}(b).
The uniqueness found in this partitioning is due to the $R$ and $E$ clusters being broken up and a few nodes being moved around.
The largest and most persistent in the bootstrapped networks cluster exactly resembles the $D$ cluster from the full network.
The second largest cluster from the full network, $R$, is broken into two clusters, each of which contains also several $E$ nodes. 
However, while the larger $R$-like cluster is highly persistent during bootstrapping (except for the $E_{05}$ node that tends to be absorbed into $E$ clusters), the smaller $R$-like cluster is much less frequent in the bootstrapped networks. 
Rather, it tends to be grouped with other $R$ nodes, specifically $R_{01}$, $R_{05}$, and $R_{08}$.
The three $F$ nodes with the addition of $E_{06}$ form the fourth cluster in this partitioning, though the latter node tends to move between clusters of the $F$ nodes and clusters of $E$ nodes during bootstrapping.
The remaining $E$ nodes form the fifth cluster, though two out of the five nodes in this group are occasionally clustered with the $F$ nodes.

%%%%%%%%%%%%%%%%%%%%%%%%%%%%%%%%%%%%%%%%%%%%
\begin{figure*}[ht!]
    \centering
    \includegraphics[width=0.96\textwidth]{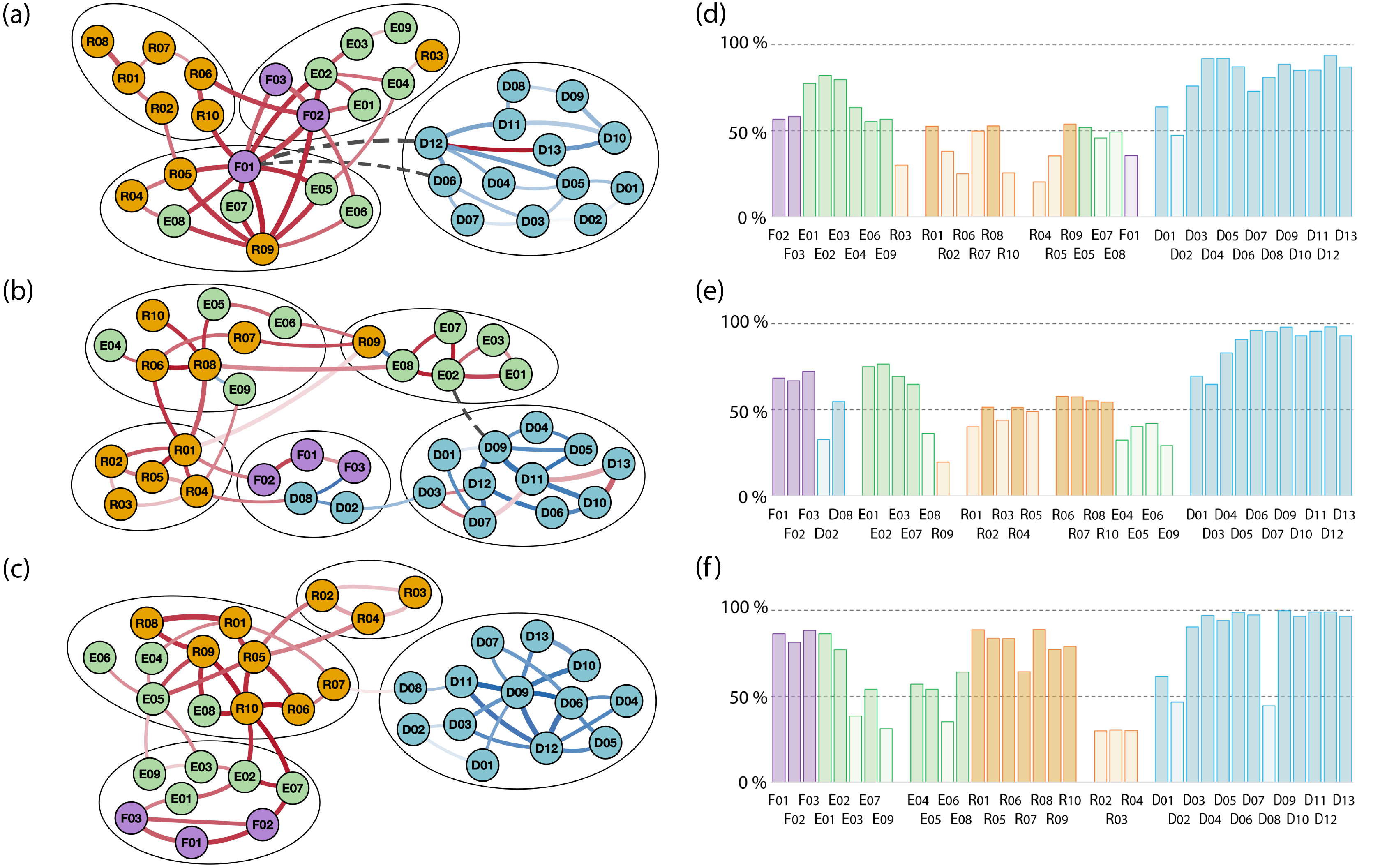}
    \caption{Plots and histograms for investigation by respondent's indicated funding source: the networks created based on the respondents receiving (a) research, (b) nonresearch, and (c) mixed funding; the frequency plots for the (d) research, (e) nonresearch, and (f) mixed networks bootstrapping.
    In the network plots, the four original themes are highlighted with different colors.
    Positive temperature edges are indicated in red and negative temperature edges are indicated in blue.
    New clusters are indicated by the circled regions.
    Each histogram is grouped by the clusters in the respective network, with nodes not reaching 50~\% stability made transparent.
    }
    \label{fig:networks-support}
\end{figure*}
%%%%%%%%%%%%%%%%%%%%%%%%%%%%%%%%%%%%%%%%%%%%

%%%%%%%%%%%%%%%%%%%%%%%%%%%%%%%%%%%%%%%%%%%%%%%%%%%%%%%%%%%%%%
\subsection{Exploring experiences between early and later semester respondents}
%%%%%%%%%%%%%%%%%%%%%%%%%%%%%%%%%%%%%%%%%%%%
The next comparison we consider is between students in their early semesters (four or fewer) and respondents in their later semesters (five or more) of graduate school.
The network that results from early semester respondents (henceforth known as the {\it early network}) is partitioned into three clusters, as seen in Fig.~\ref{fig:networks-semester}(a).
The network that results from later semester respondents (the {\it later network}) is partitioned into four different clusters, as seen in Fig.~\ref{fig:networks-semester}(b).

The NDC similarity is again very high for all three comparisons, as seen in Table~\ref{tab:mod_meas}.
The EEJ similarity varies significantly between the pairwise comparisons, with the early to full network $\mathrm{EEJ}=0.39$ and the later to full network $\mathrm{EEJ}=0.73$.
The structures of the two program-based networks when compared side by side are even more inconsistent, with $\mathrm{EEJ}=0.29$.
Interestingly, the purity is somewhat higher for the early-to-full network comparison than for the later-to-full network, indicating that although the specific edges of the network are highly varying for the early semester respondents, they have a more similar clustering structure.

The early network is made up of three large clusters, depicted in Fig.~\ref{fig:networks-semester}(a), all of which are very stable over the sampling process, see Fig.~\ref{fig:networks-semester}(c).
The first cluster recreates the $D$ structure except for the $D_{02}$ node which tends to move between clusters.
The second cluster consists of all of the $R$ nodes and three $E$ nodes ($E_{04}$, $E_{05}$, and $E_{08}$), with $E_{04}$ and $R_{03}$ occasionally moving together to the third, $F+E$ cluster.
The third cluster includes all of the $F$ nodes and all of the remaining $E$ nodes, with the $F_{02}$ playing a central role in connecting other nodes.

The partitioning of the later network includes five clusters, as depicted in Fig.~\ref{fig:networks-semester}(b).
The $D$ and $F$ clusters are held together while the $R$ and $E$ clusters are split into three clusters: an $R$-like cluster consisting of seven $R$ nodes and the $E_{04}$ node (which tends to fall out of this cluster a majority of the time); an $R+E$ cluster consisting of four $E$ nodes and three $R$ nodes; and an $E$ cluster consisting of the remaining four $E$ nodes.
The persistence of the mixed $E+R$ cluster,  as seen in Fig.~\ref{fig:networks-semester}(d), is one of the most unique features of the later semester network.
The second unique feature is the lack of persistence of the $F$ cluster which tends to get pulled into larger $R+E$ clusters through connections with $E_{01}$ and $E_{02}$.
Finally, the smaller group of $E$ nodes is fairly consistent as well but often groups up with the other $E$ nodes.
\vspace{-10pt}

%%%%%%%%%%%%%%%%%%%%%%%%%%%%%%%%%%%%%%%%%%%%%%%%%%%%%%%%%%%%%%
\subsection{Exploring experiences based on available financial support}
%%%%%%%%%%%%%%%%%%%%%%%%%%%%%%%%%%%%%%%%%%%%
The final split of the network we consider is based on the type of funding that students reported receiving during their time in graduate school.
The network built from the respondents who reported relying exclusively on research or fellowship funding during their time in graduate school (the {\it research network}) is partitioned into four clusters, as seen in Fig~\ref{fig:networks-support}(a).
The network built from the respondents who reported only relying on teaching assistantships, or other sources of funding during their time in graduate school (the {\it nonresearch network}) is partitioned into five clusters, as seen in Fig~\ref{fig:networks-support}(b).
The final network in this comparison is built from respondents who reported a mixture of research-based on nonresearch-based funding sources (the {\it mixed network}).
The mixed network is partitioned into four clusters in our analysis.

Similar to the past sections, we make direct comparisons based on node degree and the edges that build the network, as seen in Table~\ref{tab:mod_meas}.
Of these three support-based networks, the mixed network is most similar to the full network, with $\mathrm{NDC}$=$0.96$ and $\mathrm{EEJ}$=$0.59$.
While the NDC remains fairly high for the other two comparisons, the EEJ falls below 0.5, indicating that less than $50~\%$ of edges are the same between the research and nonresearch networks when compared to the full network.
The partitioning purity level is consistent with the other three demographic-based splits at around $0.7$ to $0.8$.
Interestingly, the research and nonresearch networks are the most dissimilar out of all compared networks both structurally and in terms of partitioning, with $\mathrm{EEJ}$=$0.19$ \& purity $0.61$.

The results of the sampling process for all three networks are depicted in Figs.~\ref{fig:networks-support}(d)--~\ref{fig:networks-support}(f).
In all three networks, the most persistent is the $D$ cluster with the exception of the $D_{02}$ node (in the research and mixed networks) and the $D_{08}$ node (in the mixed network). 
In the research and mixed networks, the unstable $D_{02}$ tends to move around between the other clusters.
In the nonresearch network, these two unstable $D$ nodes tend to get grouped with the $F$ nodes, forming an $F+D$ cluster, though only the $D_{08}$ node remains in this cluster throughout the sampling process while $D_{02}$ often gets pulled back into the $D$ cluster.

Two out of the three $F$ nodes are grouped with a subset of $E$ nodes in the research network, forming a persistent $R+E+F$ cluster.
However, the single $R$ node, also originally assigned to this cluster, tends to move into clusters with the other $R$ nodes.
The third $F$ node, originally assigned to the smaller and less stable $R+E+F$ cluster, tends to get reabsorbed by the bigger and more stable $R+E+F$ cluster. 
The research network has some very interesting features, including the node  $F_{01}$ playing a central role in connecting all clusters with positive temperature edges while also being connected to two $D$ nodes through dissimilar edges and a lack of persistent $R$ cluster.

The third, bigger $R+E$ cluster in the nonresearch network consists of four $R$ nodes (persistently assigned to the same cluster) and four $E$ nodes that are sometimes grouped into the fourth, smaller $E+R$ cluster, and sometimes move around between other clusters. 
Finally, the fifth $R$ cluster in the nonresearch network is only about half of the time grouped on its own, while the other half is grouped back with the other $R$ nodes.

Out of the three remaining clusters in the mixed network, the largest $E+R$ one consists of seven $R$ nodes, all of which persist during the bootstrapping tests, and four $E$ nodes, out of which one ($E_{06}$) tends to be reabsorbed by the fourth, $F+E$ cluster. 
The $F+E$ cluster consists of all of the $F$ nodes (persisting throughout the bootstrapping tests) and five $E$ nodes (two of which tend to frequently get grouped with the $E+R$ cluster).
Finally, the smallest three-node $R$ cluster does not persist throughout the sampling process and instead is reabsorbed by the $E+R$ cluster.

%%%%%%%%%%%%%%%%%%%%%%%%%%%%%%%%%%%%%%%%%%%%%%%%%%%%%%%%%%%%%%
\section{Interpretation and discussion}
\label{sec:inter}
%%%%%%%%%%%%%%%%%%%%%%%%%%%%%%%%%%%%%%%%%%%%%%%%%%%%%%%%%%%%%%
In this work, we focus on three questions related to the utility of the NALS methodology to study the ASES dataset. 
The first research question that guides this study relates to the stability of the NALS clustering.
When performing bootstrapping on the full dataset, we find that all of the thematic clusters identified by~\citet{Dalka22-NALS} pass the threshold test, though the persistence level varies between the clusters from $0.95(6)$ for the professional and academic development theme (cluster $D$) to $0.63(0)$ for the financial support theme (cluster $F$).
This confirms that the NALS methodology produces stable thematic clusters from the full dataset.

When considering demographic-based networks, we find that for some demographic groups, NALS produces several new clusters that typically include nodes from multiple original clusters. 
However, some of the new clusters, especially the smaller ones, turn out to be unstable.
Rather, they tend to revert to the original $R$, $E$, $D$, and $F$ clusters.
However, we also see that a certain level of divergence from the full network persists in the demographic-based networks, indicating that there are aspects of experiences that are unique to specific demographic groups.
Thus, answering the second research question, the clusters do change for well-defined respondent groups and while not all changes are meaningfully significant, there are a few strong changes due to shifts of small groups of nodes within particular partitionings.

These differences between groups of respondents can be further explored to identify potentially unique emergent themes and help us answer the third question pertaining to differences in student experiences. 
In this section, we will look through each of the original thematic clusters and how they are represented within the demographic-based networks.

\begin{center}
\mbox{\parbox{0.9\linewidth}{Observation 1: A majority of students do not experience professional and academic development, but network features highlight a variability within this theme.\strut}}
\end{center}

Out of the four original thematic clusters, the professional and academic development theme (cluster $D$) is most often recovered in the demographic-based networks.
In seven out of the ten considered networks, the $D$ nodes are primarily connected through negative temperature similarity.
For the remaining three networks, the positive temperature edges connect only a small subset of nodes: a single edge connecting mentoring ($D_{11}$) and PI ($D_{12}$) training in women network, a single edge connecting PI ($D_{12}$) and networking ($D_{13}$) training in the research network, and a path connecting pairwise a subset of nodes related to the various aspects of professional and academic training in the nonresearch funding network [PI training ($D_{12}$), structured collaboration ($D_{03}$), tutoring ($D_{07}$), mentoring training ($D_{11}$), networking training ($D_{13}$), and career training ($D_{10}$)].
This indicates that women and students with exclusively research or nonresearch support do experience the aspects of professional and academic development identified above.
However, the significant prevalence of negative temperature edges for the professional and academic development theme suggests that the majority of students, regardless of their demographics, are not experiencing these supports within their department.

In addition to the positive temperature edges, there are several other interesting features in the $D$ cluster for certain demographic-based networks.
For example, in the nonbridge network, we see that the small three-node cluster of $D_{01}$, $D_{02}$, and $E_{09}$ persists throughout the bootstrapping tests.
The first two items are most strongly associated with the coursework part of the professional and academic development theme corresponding to academic assessment ($D_{01}$) and personalization ($D_{02}$).
Persistent grouping of those two items with the coursework support ($E_{09}$) through negative temperature edges suggests that for the nonbridge respondents in our sample, the coursework-related supports are missing in unique ways compared to the other types of development supports.
This result is consistent with Sachmpazidi and Henderson's finding that students in bridge-affiliated programs report better social and academic integration~\cite{Sachmpazidi21-DSS}.
However, the results of this study allow us to identify the specific items that this applies to and investigate their connections to the other items within the instrument.
Additionally, it calls us to question how the naming of the original cluster from the full dataset informs our interpretation.
Here, we see how the coursework support that would be a part of the ``academic development'' aspect of the $D$ cluster is not necessarily a universally similar experienced type of support to the ``professional development'' aspect of the $D$ cluster.

Looking at the types of connections that the $D$ nodes have extending outside of that thematic cluster allows for additional interpretation.
For example, in the research network, the first thing that stands out are the two edges of dissimilarity connecting tuition ($F_{01}$) to both time management training and PI training ($D_{06}$ and $D_{12}$, respectively).
This means that while graduate students supported solely through research assistantships have their tuition fully supported (all other edges connecting this node within the network have a positive temperature), this aspect of their experiences is highly correlated with a lack of training provided by their institution in how to run a lab or manage their time.
In other words, although these students may not worry about their financial stability in graduate school, they may not be supported in advancing certain skills important for their academic careers.

\begin{center}
\mbox{\parbox{0.9\linewidth}{Observation 2: Financial supports are widely experienced and are connected with social and scholarly exploration for students in departments with bridge programs and early semester students.\strut}}
\end{center}

The second theme persistent in all but one considered network is the financial support (cluster $F$).
The three $F$ nodes are almost always clustered together within the demographic-based networks, however, they often tend to get grouped into larger clusters with additional nodes from other themes.
While these larger clusters are not always fully stable, we do find that, in some cases, most of the non-$F$ nodes persist throughout the bootstrapping tests.
One such example is the bridge network.
Here, we have the three $F$ nodes (tuition, health, and life) clustered at a stable level with three nodes from the social and scholarly exploration theme (cluster $E$), including socializing, shared space, and research survey.
This suggests that for respondents in the bridge program, the financial support is experienced in similar patterns with support around building community and understanding the research available.
This indicates that departments with bridge programs in addition to financial support offer a more holistic support that includes, e.g., space for socializing and research exploration.
The two unstable nodes, research exploration ($E_{06}$) and flexibility ($E_{08}$), further support this observation.
The mixed funding support network has the same combination of stable nodes in the cluster as the bridge network.

The same type of cluster (i.e., $F$ + $E$) is also seen in the early semesters network, though here two additional nodes---accommodations ($E_{03}$) and research exploration ($E_{06}$)---are being held stable in the cluster.
From this, we can conclude that for these particular groups of students---students in departments with bridge programs, students who have received funding from multiple sources, and students who are early in their graduate studies---the financial support they receive is experienced in similar ways to the supports for social and scholarly exploration.
These experiences are based on the support that the departments are offering.
Thus, as expected, we see that bridge programs and students early in graduate studies have different types of support available to and designed for them.

The only network in which the $F$ nodes are separated is the research network, where the central position played by the tuition item ($F_{01}$) helps to mix around the clustering in the sampled networks.
In fact, it is the only one in which an $F$ node is the most central in terms of both the local and global connectivity, with $\mathcal{C}_D(F_{01}) = 11$ and $\mathcal{C}_B(F_{01}) = 348$, respectively, which is an artifact of little variability in the ways research-only supported students answered this question.
This shows how one strong experience shared by a majority of respondents can influence the clustering of the rest of the themes in the NALS approach.

\begin{center}
\mbox{\parbox{0.9\linewidth}{Observation 3: Social and scholarly exploration experiences are highly variable and depend on the student group.\strut}}
\end{center}

The $E$ clusters in the demographic-based networks are some of the most variable.
Sometimes they form clusters with $F$ nodes, as described above.
In other networks, they form smaller clusters on their own or pair up with $R$ nodes to form larger clusters.
For example, $E$ nodes that clustered together with the $F$ nodes to form a larger cluster in the mixed support network (i.e., socializing, shared space, and research survey), in the nonresearch support network form their own stable cluster (along with accommodations).
We can thus conclude that these aspects of social and scholarly support are uniquely experienced compared to other supports by students who have been supported solely by teaching assistantships or other forms of funding.
It may also be that institutions with less established funding pathways have fewer support options for students socially, for research surveys, and for getting accommodations.

In previous work, peer mentoring has been identified as a central support structure that could be made more formalized to better serve students~\cite{Sachmpazidi21-RDS}.
In our analysis, we see how the peer mentorship item ($E_{04}$) is often weakly clustered with nodes from different themes, spending time in both $R$ clusters and $E$ clusters.
The only two demographic group networks where this item is stably clustered are the research network---where it is grouped with other social-centric items---and the mixed network---where it is grouped with mainly research-centric items.
The peer mentorship that has been identified as important to build up may need to be connected up with different types of experiences for different students.

Additionally, we can see how some of the other $E$ nodes are occasionally clustered with $R$ nodes, such as in the later semesters network.
This cluster brings together the socializing, shared space, research survey, and research flexibility (nodes $E_{01}$, $E_{02}$, $E_{07}$, and $E_{08}$, respectively) with meetings consistency, project matching, and presentations (nodes $R_{06}$, $R_{09}$, and $R_{10}$, respectively). 
One could imagine several reasons why these types of support would be experienced in similar ways by graduate students later on in their careers.
It could be that for later career graduate students these supports have been well established and remain consistent as they continue their work.
Since by this time in graduate school, most students will have found a permanent research group or topic of study that they are committed to, they perceive the research survey and flexibility as integral aspects of experiences that led to project matching or defining meeting consistency.

From a methodological point of view, our analysis of the $E$ cluster revealed the flaws in assuming that an item grouping in cluster analysis is relevant across all populations.
Thus, when naming clusters in an attempt to capture a dynamic set of experiences, caution should be taken to not place too much emphasis on the names of the themes.
In this work, we used names that most closely resemble names used in the original ASES work~\cite{Sachmpazidi21-DSS} to make it easier for the readers familiar with the ASES work to connect clusters derived using NALS with the original ones.
We have also shown that in analyzing the network of survey items, researchers can identify ways that these themes are built and how particular groupings are more or less stable across different populations.
Such new themes and differences in thematic clusters revealed through the NALS approach can lead to a more in-depth analysis through more qualitative methods.

In previous qualitative work, social and academic supports were identified as some of the most influential aspects of graduate student persistence~\cite{Sachmpazidi22-CEW}.
Men graduate students in bridge programs were found to have experienced a social support system that was a deliberate part of their program, while women in nonbridge programs experienced an explicit lack of it.
In the bridge network, we see that the $E$ nodes related to social support are more connected to the other identified thematic nodes than in the nonbridge network.
In the women network, the clustering of $E$ nodes is highly unstable.
Our analysis offers additional context for previous findings in that the social support structures are more well connected to other forms of support and stable for bridge students, while they are experienced in highly variable ways for women in the dataset.

\begin{center}
\mbox{\parbox{0.9\linewidth}{Observation 4: Mentoring and research experiences are formed from subthemes that highlight different connections for different student groups.\strut}}
\end{center}

The mentoring and research experience thematic $R$ cluster is well represented in most of the demographic-based networks.
Except for a few individual $R$ nodes moving around, the core nodes of the cluster stay together, such as in the bridge and early semesters networks.
This means that for many of the demographic groups of students represented in this survey, the mentoring and research experience supports are experienced together.
Interestingly, for the nonbridge and the later semesters networks, the same three $R$ nodes---meetings consistency, project matching, and presentations---get pulled out, leaving the $R$ thematic cluster focused mainly on interactions with research mentors.
This suggests that for these groups of students, the relationship with their research mentor is highly important.

Another pattern we see within the $R$ thematic cluster is the emergence of ``minithemes'' that hold together a small number of nodes during the sampling process.
While these small clusters are not necessarily seen in the overall partitioning, they appear during the sampling.
For example, in the women network, although the larger clusters involving $R$ nodes are unstable, there are some small groups of nodes within these clusters that consistently cluster together in the sampled networks.
One such group, related to presenting and discussing research progress and consisting of research meetings, regular feedback, and presentations (nodes $R_{01}$, $R_{08}$, and $R_{10}$, respectively), emerges in women network.
Minithemes consisting of just two nodes that are persistently grouped together are research meetings and regular feedback in the research network and academic planning and integration in the nonresearch network.

When researchers assume thematic clusters hold true across all student populations, there is a danger in inflating the importance of the theme names, rather than the specific experiences that make up that theme.
Through using the NALS methodology on data representing demographic groups, we are able to investigate whether the ASES themes are represented within each of these groups and whether any new themes emerge for them.
We find that while some thematic clusters from the original network are well represented in the demographic-based networks, other clusters behave in different and unique ways, confirming that experiences are different for each of the demographic groups.
Although the new clusters are not always stable, there are particular features that are significant within the networks that help us shed light on the unique needs of support structures for different demographic groups of graduate students.

%%%%%%%%%%%%%%%%%%%%%%%%%%%%%%%%%%%%%%%%%%%%%%%%%%%%%%%%%%%%%%
\section{Conclusions}\label{sec:conclusion}
%%%%%%%%%%%%%%%%%%%%%%%%%%%%%%%%%%%%%%%%%%%%%%%%%%%%%%%%%%%%%%
In this paper, we have focused on two aspects of ASES data analysis using the NALS methodology: the thematic variability between various demographic-based networks and the stability of the resulting themes.
Using the full ASES dataset, we have confirmed that the thematic clusters found through NALS are stable against small data perturbation which indicates that the themes are well suited to capture patterns in graduate students' experiences.
Additionally, we have shown that for demographic-based networks, NALS can reveal certain unique features that shed light on the needs of particular demographic groups.

For each of the four thematic clusters, we observed varying trends across the demographic-based networks.
For example, not experiencing professional and academic development was a shared theme among many of the groups considered in this work, indicating that regardless of demographics, students generally lack formal support from their programs in this domain.
Several groups of students---students in departments with bridge programs, students who have received funding from multiple sources, and students who are early in their graduate studies---reported experiencing financial support in similar ways as certain aspects of social and scholarly exploration.
For women, on the other hand, the financial support was a unique enough experience to form a persistent, stand-alone theme while for students with nonresearch support, who typically rely on teaching assistantship throughout graduate school, financial support experiences were strongly connected with teaching training. 

Additionally, for some demographic-based networks---like students in later semesters of their graduate studies---themes tend to mix social and scholarly exploration with mentoring and research experience, indicating that the early exploration may tie strongly to their later research experiences for particular groups.
Finally, we are able to identify minithemes, groupings of two or more items that cluster together strongly. We found these appearing most in the mentoring and research experience theme.

Through the use of the NALS methodology, we have identified unique features related to each ASES theme for different demographic groups.
These features, made explicit by a network approach, open new opportunities for interpretation of the survey data.
They indicate areas for future research into the connections between the various support structures as experienced by different demographic groups.
These results can inform deliberately designed support structures targeting distinct groups of graduate students within physics departments.
\vspace{-10pt}

%%%%%%%%%%%%%%%%%%%%%%%%%%%%%%%%%%%%%%%%%%%%%%%%%%%%%%%%%%%%%%
\subsection{Limitations and future work}
%%%%%%%%%%%%%%%%%%%%%%%%%%%%%%%%%%%%%%%%%%%%%%
We present the limitations of this work in two parts: limitations due to the dataset and limitations due to the methodology.
We also discuss how future work could address these limitations and make further recommendations for improving graduate physics programs.

The ASES dataset represents only a small fraction of physics graduate students nationwide.
Additionally, the fluctuation in response rates between departments may result in overrepresentation of some departments which, in turn, might influence particular demographic responses.
Some demographic groups, such as nonbinary students or groups separated by specific racial and ethnic backgrounds, were strongly underrepresented in the ASES dataset and thus we had to omit them in our analysis.
We urge researchers across various physics graduate programs to use the ASES instrument to build a more complete picture of the landscape of experiences related to support structures.

Previously, ASES has been shown to reliably measure differences between student experiences within bridge and nonbridge programs by comparing the quantitative differences of a principle component analysis with qualitative analysis of interviews about those programs' support structures~\cite{Sachmpazidi21-DSS}.
However, evidence of measurement invariance across other variables of respondent characteristics was not included.
For example, it is possible that students who are later in their graduate school careers would interpret particular survey items differently than students in earlier stages of their programs.
Thus, these two student groups may report different experiences due to differences in interpretation of items rather than differences in experiences.
While there is evidence of measurement invariance between bridge and nonbridge programs, the lack of evidence across other demographic characteristics is a limitation of this study.

Additionally, our dataset only represents the experience of students at one point in time.
It would be useful to establish multiple data points in time to further explain how these support structures can relate to student outcomes and program general attrition rates.
Research should continue to grow in this area to understand the longitudinal effects of different types of support structures.
NALS would be well positioned to investigate the evolution of these networks of experiences of support structures.

Analyzing the thematic clusters has helped us identify important groups of experiences for different demographic groups.
However, survey items that are particularly strong within the network (as measured by node centrality) can influence the clustering in ways that result in less stable clusters.
This is an inherent feature of the method that should be noted as NALS is further developed.
In our analysis, we only looked at single-level demographics, however, the ways in which particular categories intersect at a broad scale may provide additional insight into the formation of the thematic clusters.
Finally, in making inferences about why particular patterns emerge, we do not have deeper insights into the experiences that could be captured by qualitative studies.
Pairing the NALS methodology with qualitative approaches when analyzing other survey datasets will likely produce more robust and comprehensive results.

%%%%%%%%%%%%%%%%%%%%%%%%%%%%%%%%%%%%%%%%%%%%
\begin{acknowledgments}
We appreciate the useful discussions, and the data-sharing collaboration, with D.~Sachmpazidi and C.~Henderson.
R. P. D. was supported by the National Science Foundation Graduate Research Fellowship under Grant No. DGE 1840340.
The views and conclusions contained in this paper are those of the authors and should not be interpreted as representing the official policies, either expressed or implied, of the National Science Foundation, or the U.S. Government.
The U.S. Government is authorized to reproduce and distribute reprints for Government purposes notwithstanding any copyright noted herein. 
Any mention of equipment, instruments, software, or materials; it does not imply recommendation or endorsement by the National Institute of Standards and Technology.
\end{acknowledgments}

%%%%%%%%%%%%%%%%%%%%%%%%%%%%%%%%%%%%%%%%%%%%%
\appendix
%%%%%%%%%%%%%%%%%%%%%%%%%%%%%%%%%%%%%%%%%%%%%
\vspace{-5pt}

\section{CLUSTERING CONVERGENCE ANALYSIS}
\label{app:convergence}
\begin{figure}[t]
    \centering
    \includegraphics[width=0.48\textwidth]{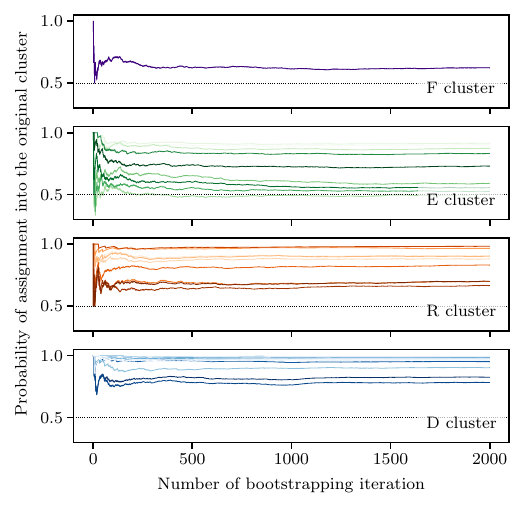}
    \caption{Clustering convergence analysis for the bootstrapped networks compared to the four thematic clusters from the full ASES dataset.
    The three nodes from cluster F have identical convergence curves.
    The assignment into the original clusters convergences at $\nu_{\mathrm F}=628$, $\nu_{\mathrm E}=705$, $\nu_{\mathrm R}=612$, and $\nu_{\mathrm D}
    =406$ for clusters F, E, R, and D, respectively.}
    \label{fig:clustering_convergence}
\end{figure}

%%%%%%%%%%%%%%%%%%%%%%%%%%%%%%%%%%%%%%%%%%%%
To determine the appropriate number of bootstrapping iterations for our dataset, we perform a clustering convergence analysis using the full ASES dataset. 
Following the procedure described in Sec.~\ref{ssec:stat-viz}, we ran $2,000$ bootstrapping tests.
We then recorded the frequencies of assigning nodes in the bootstrapped networks to their clusters from the original NALS thematic partitioning of the full backbone network.
Figure~\ref{fig:clustering_convergence} shows the convergence plots for the four thematic clusters as a function of the number of bootstrapping iterations.
While the overall frequencies vary between clusters as well as for nodes within each cluster (except for cluster F), in all cases, we observe a full convergence at around $700$ iterations. 

%%%%%%%%%%%%%%%%%%%%%%%%%%%%%%%%%%%%%%%%%%%%
%%%%%%%%%%%%%%%%%%%%%%%%%%%%%%%%%%%%%%%%%%%%
\vspace{-5pt}

%%%%%%%%%%%%%%%%%%%%%%%%%%%%%%%%%%%%%%%%%%%%
\section{RELATIONSHIP BETWEEN METRICS AND SAMPLE SIZE}
\label{app:sample-size}
%%%%%%%%%%%%%%%%%%%%%%%%%%%%%%%%%%%%%%%%%%%%
Certain demographic groups in our analysis have somewhat unequal sample sizes between the subgroups (e.g., gender-based and the number of semesters since enrollment-based splits).
Thus, we found it is important to investigate whether there exists a generalized relationship between network comparison metrics and sample size.
To do this, we ran $2,000$ bootstrapping tests at the following sample sizes: $50$, $100$, $150$, $200$, $250$, $300$, $350$, and $381$.
We then ran the NDC, EEJ, and purity similarity tests for each bootstrapped test and calculated the mean and standard deviation for each set of sample sizes.

\begin{figure}[t]
    \centering
    \includegraphics[width=0.48\textwidth]{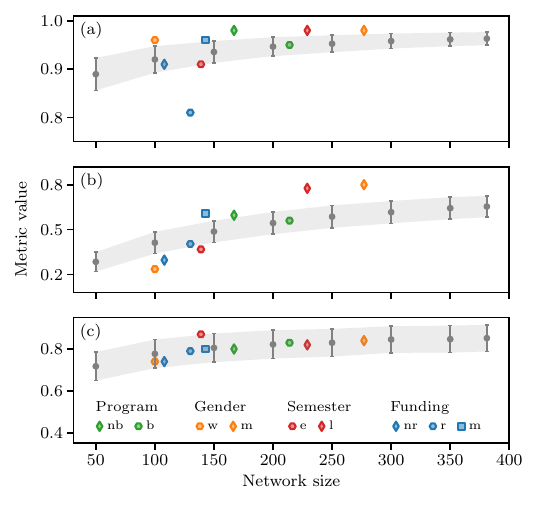}
    \caption{Network comparison metric values for different sample size bootstrapped trial networks and demographic-based networks: (a) the NDC metric value, (b) the EEJ metric value, and (c) the purity value. 
    The point shape represents different subgroups within each demographic split as shown in the legend. 
    Note the different scales on the $y$ axis for each plot.
    Error bars indicate one standard deviation.}
    \label{fig:sample_size}
\end{figure}

As one may expect, the more of the dataset that was drawn from, the more similar those networks became with respect to the full network.
The gray plot in  Fig.~\ref{fig:sample_size} confirms this trend.
In addition, Fig.~\ref{fig:sample_size} shows the NDC, EEJ, and purity values for all demographic-based networks considered in this work.
As one can see, not all of the differences can be explained simply by sample size alone.
For example, the NDC value for the midsize research-based network is significantly lower than for the other two funding networks, see Fig.~~\ref{fig:sample_size}.
For the program-based networks, both NDC [Fig.~~\ref{fig:sample_size}(a)] and EEJ [Fig.~\ref{fig:sample_size}(b)] are significantly higher for the smaller bridge network than for the nonbridge networks.
For both gender-based and semester-based networks, the EEJ values are unusually low for the smaller subgroups and unusually high for the bigger subgroups. 
Purity for all networks seems to follow the expected trend.

%%%%%%%%%%%%%%%%%%%%%%%%%%%%%%%%%%%%%%%%%%%%
%%%%%%%%%%%%%%%%%%%%%%%%%%%%%%%%%%%%%%%%%%%%
\vspace{-5pt}

%%%%%%%%%%%%%%%%%%%%%%%%%%%%%%%%%%%%%%%%%%%%
\section{THE EFFECT SIZE FOR NETWORK COMPARISON}
\label{app:effect_size}
%%%%%%%%%%%%%%%%%%%%%%%%%%%%%%%%%%%%%%%%%%%%
Cohen's $d$ measure, given in Eq.~(\ref{eq:cohen_d}) quantifies the strength of differences of network measures between the relevant bootstrapped demographic-based networks and full network.
In our analysis, we used Cohen's $d$ to compare the NDC, EEJ, and purity values based on $N=1,000$ samples.
Table~\ref{tab:cohens_d} shows the Cohen's $d$ values for all comparisons.

For the EEJ measure, all comparisons have a large effect size.
In contrast, for NDC, all but one comparison have a small effect size.
The one comparison with a medium effect size---research vs full and mixed vs full---just barely meets the $d > 0.5$ threshold.
When looking at purity, the split based on gender and the split based on semesters show large effect sizes.
The purity of research vs full has a medium effect size when compared to both nonresearch vs full and mixed vs full.

\vspace{-10pt}

%%%%%%%%%%%%%%%%%%%%%%%%%%%%%%%%%%%%%%%%%%%%%%%%%%%%%%%%%%%
\section{CENTRALITY MEASURES}
\label{app:centralities}
%%%%%%%%%%%%%%%%%%%%%%%%%%%%%%%%%%%%%%%%%%%%%%%%%%%%%%%%%%%%%%
Table~\ref{tab:cent_meas} shows the degree and betweenness centrality measures of all networks.
The degree centrality $\mathcal{C}_D$ distribution is fairly consistent between networks, ranging from $1$ to $9$, with an overall $M_{\mathcal{C}_D}=3.1(1.6)$. 
The betweenness centrality $\mathcal{C}_B$, on the other hand, varies significantly between networks in terms of both the overall magnitude and which nodes are identified as most central.
It is important to note that for networks with a two-component network, the maximum possible betweenness will be less than that of a single-component network, as more nodes are reachable by the paths.

The bridge network has many nodes with high betweenness, but the node with by far the largest is $D_{02}$ [$\mathcal{C}_B(D_{02}) = 267$].
$R_{09}$ has the highest degree in the bridge network [$\mathcal{C}_D(R_{09}) = 8$].
In the nonbridge network, $R_{01}$ and $R_{06}$ have the highest betweenness [$\mathcal{C}_B(R_{01}) = 56$ and $\mathcal{C}_B(R_{06}) = 51$] while $D_{09}$ having the largest degree [$\mathcal{C}_D(D_{09}) = 8$].
For the women network, the node with high betweenness is $D_{12}$ [$\mathcal{C}_B(D_{12}) = 182$].
The two nodes with the highest degree centrality is also $D_{12}$ [$\mathcal{C}_D(D_{12}) = 7$].
In the men network, on the other hand, $E_{05}$ has the highest betweenness [$\mathcal{C}_B(E_{05}) = 91$] and $D_{09}$ has the highest degree [$\mathcal{C}_D(D_{09}) = 8$], similar to the full network.
In the early network, $R_{09}$ has both the highest betweenness [$\mathcal{C}_B(R_{09}) = 108$] and degree [$\mathcal{C}_D(R_{09}) = 9$].
In the later network, $R_{05}$ and $R_{09}$ have the highest betweenness 
[$\mathcal{C}_B(R_{05}) = 82$, $\mathcal{C}_B(R_{09}) = 80$] while $D_{09}$ has the highest degree [$\mathcal{C}_D(D_{09}) = 8$].
In the research network, $F_{01}$ has both the highest betweenness 
[$\mathcal{C}_B(F_{01}) = 348$] and the highest degree [$\mathcal{C}_D(F_{01}) = 11$].
In the nonresearch network, $E_{02}$ has the highest betweenness [$\mathcal{C}_B(D_{08}) = 213$] while $R_{01}$ has the highest degree [$\mathcal{C}_D(R_{01}) = 7$].
Finally, $R_{07}$ has the highest betweenness [$\mathcal{C}_B(R_{07}) = 274$] and $D_{09}$ has the highest degree [$\mathcal{C}_D(D_{09}) = 8$] in the mixed network.

%%%%%%%%%%%%%%%%%%%%%%%%%%%%%%%%%%%%%%%%%%%%
\begin{table}[!th]
\renewcommand{\arraystretch}{1.02}
\renewcommand{\tabcolsep}{2pt}
\caption{\label{tab:cohens_d}
Cohen's $d$ measure for comparisons based on NDC, EEJ, and purity.
The comparisons are made between the network indicated in column 1 and the full network and the network indicated in column 2 and the full network.
A double asterisk indicates large effect $d$ values while a single asterisk indicates medium effect $d$ values.}
\begin{ruledtabular}
\begin{tabular}{lllll}
Comparison 1 & Comparison 2  & $d_{\text{NDC}}$ & $d_{\text{EEJ}}$ & $d_{\text{purity}}$ \\ \hline
Bridge      & Nonbridge
& 0.44 & $0.97^{**}$ & 0.14 \\ 
Women  & Men         
& 0.11 & $3.68^{**}$ & $1.92^{**}$ \\ 
Early       & Later   
& 0.14 & $3.14^{**}$ & $1.16^{**}$ \\ 
Research    & Nonresearch
& 0.49 & $1.75^{**}$ & $0.54^{*}$ \\ 
Research    & Mixed
& $0.50^{*}$ & $1.18^{**}$ & $0.64^{*}$ \\ 
Nonresearch & Mixed
& 0.01 & $3.02^{**}$ & 0.06 \\ 
\end{tabular}
\end{ruledtabular}
\end{table}
%%%%%%%%%%%%%%%%%%%%%%%%%%%%%%%%%%%%%%%%%%%%

\vspace{-15pt}

%%%%%%%%%%%%%%%%%%%%%%%%%%%%%%%%%%%%%%%%%%%%%%%%%%%%%%%%%%%
\section{NALS THEMES OF THE ASES INSTRUMENT}
\label{app:nals_themes}
%%%%%%%%%%%%%%%%%%%%%%%%%%%%%%%%%%%%%%%%%%%%%%%%%%%%%%%%%%%%%%
Table~\ref{tab:nals_themes} provides additional context for the thematic clustering identified through NALS~\cite{dalka2022restoring}. 
It includes the code for each survey item along with the shorthand name and the exact text that is used in the ASES instrument. 
The cluster titles are included at the beginning of each grouping.

%%%%%%%%%%%%%%%%%%%%%%%%%%%%%%%%%%%%%%%%%%%%

%%%%%%%%%%%%%%%%%%%%%%%%%%%%%%%%%%%%%%%%%%%%
\onecolumngrid

\renewcommand{\arraystretch}{1.01}
\renewcommand{\tabcolsep}{5.5pt}
\begin{longtable}[b]{c|cr|cr|cr|cr|cr|cr|cr|cr|cr|cr}
\caption{Comparison of network centralities for the full and demographic split-based networks. 
The nodes are sorted according to the themes found for the full ASES network.
Betweenness is rounded to the nearest whole number.
}
\label{tab:cent_meas} \\ \hline \hline
\multicolumn{1}{c|}{\multirow{3}{*}{ID}} & \multicolumn{2}{c|}{\multirow{2}{*}{Full}} & \multicolumn{2}{c|}{\multirow{2}{*}{Bridge}} & \multicolumn{2}{c|}{Non-} & \multicolumn{2}{c|}{\multirow{2}{*}{Women}} & \multicolumn{2}{c|}{\multirow{2}{*}{Men}} & \multicolumn{2}{c|}{\multirow{2}{*}{Early}} & \multicolumn{2}{c|}{\multirow{2}{*}{Later}} & \multicolumn{2}{c|}{\multirow{2}{*}{Research}} & \multicolumn{2}{c|}{Non-} & \multicolumn{2}{c}{\multirow{2}{*}{Mixed}} \\
& & & & & \multicolumn{2}{c|}{bridge} & & & & & & & & & & & \multicolumn{2}{c|}{research} & \\
 & $\mathcal{C}_D$ & $\mathcal{C}_B$ & $\mathcal{C}_D$ & $\mathcal{C}_B$ & $\mathcal{C}_D$ & $\mathcal{C}_B$ & $\mathcal{C}_D$ & $\mathcal{C}_B$ & $\mathcal{C}_D$ & $\mathcal{C}_B$ & $\mathcal{C}_D$ & $\mathcal{C}_B$ & $\mathcal{C}_D$ & $\mathcal{C}_B$ & $\mathcal{C}_D$ & $\mathcal{C}_B$ & $\mathcal{C}_D$ & $\mathcal{C}_B$ & $\mathcal{C}_D$ & $\mathcal{C}_B$ \\ \hline 
\endfirsthead
\caption[]{\emph{(Continued)}} \\
\hline\hline
\multicolumn{1}{c|}{\multirow{3}{*}{ID}} & \multicolumn{2}{c|}{\multirow{2}{*}{Full}} & \multicolumn{2}{c|}{\multirow{2}{*}{Bridge}} & \multicolumn{2}{c|}{Non-} & \multicolumn{2}{c|}{\multirow{2}{*}{Women}} & \multicolumn{2}{c|}{\multirow{2}{*}{Men}} & \multicolumn{2}{c|}{\multirow{2}{*}{Early}} & \multicolumn{2}{c|}{\multirow{2}{*}{Later}} & \multicolumn{2}{c|}{\multirow{2}{*}{Research}} & \multicolumn{2}{c|}{Non-} & \multicolumn{2}{c}{\multirow{2}{*}{Mixed}} \\
& & & & & \multicolumn{2}{c|}{bridge} & & & & & & & & & & & \multicolumn{2}{c|}{research} & \\
 & $\mathcal{C}_D$ & $\mathcal{C}_B$ & $\mathcal{C}_D$ & $\mathcal{C}_B$ & $\mathcal{C}_D$ & $\mathcal{C}_B$ & $\mathcal{C}_D$ & $\mathcal{C}_B$ & $\mathcal{C}_D$ & $\mathcal{C}_B$ & $\mathcal{C}_D$ & $\mathcal{C}_B$ & $\mathcal{C}_D$ & $\mathcal{C}_B$ & $\mathcal{C}_D$ & $\mathcal{C}_B$ & $\mathcal{C}_D$ & $\mathcal{C}_B$ & $\mathcal{C}_D$ & $\mathcal{C}_B$ \\ \hline
\endhead
\hline
\multicolumn{21}{r}{\emph{(Table continued)}}
\endfoot
\endlastfoot
\hline
F01 & 2 & 0 & 3 & 3 & 2 & 0 & 2 & 0 & 2 & 0 & 2 & 0 & 4 & 38 & 11 & 348 & 2 & 10 & 2 & 0 \\
F02 & 3 & 38 & 4 & 40 & 4 & 41 & 3 & 64 & 4 & 42 & 7 & 60 & 3 & 4 & 7 & 77 & 2 & 26 & 3 & 42 \\
F03 & 2 & 0 & 3 & 5 & 2 & 0 & 2 & 0 & 2 & 0 & 3 & 7 & 2 & 0 & 2 & 0 & 2 & 15 & 3 & 4 \\ \hline
E01 & 2 & 0 & 3 & 81 & 2 & 0 & 2 & 41 & 2 & 0 & 3 & 2 & 2 & 0 & 2 & 0 & 2 & 0 & 2 & 20 \\
E02 & 5 & 24 & 4 & 76 & 6 & 41 & 4 & 141 & 3 & 12 & 6 & 67 & 4 & 21 & 5 & 104 & 5 & 213 & 4 & 62 \\
E03 & 3 & 13 & 4 & 186 & 3 & 10 & 3 & 40 & 3 & 20 & 3 & 12 & 3 & 13 & 2 & 33 & 2 & 0 & 3 & 19 \\
E04 & 2 & 9 & 1 & 0 & 1 & 0 & 2 & 35 & 2 & 7 & 3 & 41 & 2 & 6 & 3 & 35 & 1 & 0 & 2 & 20 \\
E05 & 6 & 106 & 5 & 169 & 4 & 32 & 3 & 38 & 6 & 91 & 3 & 30 & 5 & 50 & 3 & 27 & 2 & 4 & 6 & 89 \\
E06 & 2 & 54 & 2 & 13 & 2 & 16 & 2 & 11 & 2 & 37 & 2 & 0 & 2 & 7 & 2 & 0 & 2 & 2 & 1 & 0 \\
E07 & 3 & 4 & 5 & 41 & 2 & 5 & 3 & 132 & 4 & 28 & 2 & 0 & 4 & 13 & 2 & 0 & 2 & 0 & 3 & 70 \\
E08 & 3 & 26 & 2 & 0 & 3 & 19 & 2 & 35 & 2 & 23 & 2 & 60 & 2 & 6 & 3 & 14 & 4 & 198 & 2 & 0 \\
E09 & 2 & 0 & 2 & 0 & 2 & 0 & 2 & 19 & 2 & 0 & 2 & 1 & 2 & 0 & 1 & 0 & 2 & 14 & 2 & 0 \\ \hline
R01 & 5 & 14 & 5 & 18 & 5 & 56 & 4 & 76 & 3 & 12 & 3 & 0 & 6 & 31 & 3 & 37 & 7 & 149 & 4 & 107 \\
R02 & 3 & 10 & 2 & 0 & 3 & 14 & 3 & 34 & 3 & 8 & 2 & 4 & 3 & 10 & 2 & 50 & 3 & 8 & 3 & 16 \\
R03 & 2 & 0 & 2 & 33 & 2 & 0 & 3 & 139 & 2 & 0 & 2 & 7 & 2 & 0 & 1 & 0 & 2 & 3 & 2 & 0 \\
R04 & 3 & 10 & 3 & 64 & 3 & 10 & 4 & 120 & 4 & 14 & 4 & 13 & 3 & 10 & 2 & 1 & 5 & 90 & 3 & 16 \\
R05 & 5 & 59 & 6 & 143 & 4 & 29 & 4 & 139 & 6 & 79 & 2 & 4 & 6 & 82 & 4 & 90 & 3 & 3 & 6 & 148 \\
R06 & 3 & 10 & 3 & 22 & 4 & 51 & 3 & 105 & 3 & 12 & 3 & 1 & 3 & 13 & 3 & 41 & 4 & 42 & 3 & 168 \\
R07 & 2 & 3 & 2 & 1 & 2 & 3 & 2 & 137 & 2 & 1 & 2 & 0 & 1 & 0 & 2 & 15 & 2 & 2 & 3 & 274 \\
R08 & 2 & 2 & 3 & 7 & 2 & 4 & 2 & 17 & 2 & 3 & 4 & 1 & 2 & 0 & 1 & 0 & 6 & 155 & 2 & 9 \\
R09 & 6 & 75 & 8 & 140 & 5 & 27 & 4 & 84 & 5 & 52 & 9 & 108 & 6 & 80 & 7 & 35 & 4 & 87 & 4 & 22 \\
R10 & 2 & 0 & 2 & 0 & 3 & 20 & 2 & 4 & 2 & 3 & 3 & 0 & 3 & 12 & 2 & 17 & 1 & 0 & 6 & 178 \\ \hline
D01 & 2 & 4 & 3 & 84 & 2 & 0 & 2 & 0 & 3 & 5 & 3 & 4 & 2 & 4 & 2 & 3 & 2 & 5 & 2 & 10 \\
D02 & 2 & 1 & 4 & 267 & 3 & 22 & 2 & 0 & 2 & 1 & 2 & 3 & 2 & 1 & 2 & 1 & 2 & 73 & 2 & 1 \\
D03 & 3 & 6 & 2 & 2 & 3 & 30 & 3 & 64 & 3 & 7 & 2 & 0 & 3 & 8 & 4 & 33 & 3 & 72 & 3 & 22 \\
D04 & 2 & 0 & 2 & 0 & 2 & 0 & 1 & 0 & 2 & 0 & 2 & 2 & 2 & 0 & 2 & 0 & 2 & 0 & 2 & 0 \\
D05 & 3 & 3 & 5 & 39 & 3 & 1 & 2 & 0 & 4 & 4 & 2 & 1 & 4 & 4 & 4 & 40 & 3 & 2 & 2 & 0 \\
D06 & 4 & 3 & 3 & 1 & 3 & 2 & 4 & 112 & 4 & 3 & 4 & 5 & 3 & 1 & 4 & 78 & 2 & 5 & 5 & 7 \\
D07 & 2 & 0 & 4 & 150 & 2 & 0 & 3 & 138 & 2 & 0 & 4 & 15 & 2 & 0 & 2 & 0 & 4 & 32 & 2 & 0 \\
D08 & 2 & 0 & 2 & 0 & 2 & 0 & 2 & 6 & 2 & 0 & 2 & 1 & 2 & 0 & 2 & 11 & 3 & 93 & 2 & 264 \\
D09 & 9 & 41 & 6 & 13 & 8 & 37 & 6 & 70 & 8 & 32 & 7 & 32 & 8 & 35 & 2 & 1 & 6 & 184 & 8 & 161 \\
D10 & 2 & 3 & 3 & 1 & 2 & 0 & 2 & 0 & 3 & 3 & 4 & 6 & 2 & 2 & 3 & 23 & 3 & 3 & 2 & 0 \\
D11 & 4 & 8 & 7 & 94 & 3 & 1 & 3 & 9 & 4 & 9 & 3 & 7 & 5 & 11 & 3 & 66 & 5 & 66 & 3 & 253 \\
D12 & 5 & 9 & 5 & 16 & 7 & 26 & 7 & 182 & 5 & 11 & 3 & 4 & 5 & 12 & 6 & 214 & 4 & 63 & 6 & 89 \\
D13 & 2 & 1 & 2 & 0 & 2 & 0 & 2 & 0 & 2 & 1 & 2 & 0 & 2 & 1 & 2 & 25 & 2 & 0 & 2 & 0 
% %%%%%%%%%%%%%%%%%%%%%%%%%%%%%%%%%%%%%%%%%%%%
\\ \hline \hline
\end{longtable}
\twocolumngrid
%%%%%%%%%%%%%%%%%%%%%%%%%%%%%%%%%%%%%%%%%%%%

\onecolumngrid

\renewcommand{\arraystretch}{1.02}
\renewcommand{\tabcolsep}{2pt}
\begin{longtable}[H]{cp{0.18\linewidth}p{0.75\linewidth}}
\caption{NALS themes of the ASES instrument.
The first column indicates the item code, the second column gives the shorthand name for an item, and the third column gives the item's full description.}
\label{tab:nals_themes} \\ \hline \hline
ID  & Shorthand name & Full item description \\ \hline
\endfirsthead
\caption[]{\emph{(Continued)}} \\
\hline\hline
ID  & Shorthand name & Full item description \\ \hline
\endhead
\hline
\multicolumn{3}{r}{\emph{(Table continued)}}
\endfoot
\endlastfoot
\hline 
\multicolumn{3}{l}{\textit{Financial support} ($F$)} \\ \cline{1-3}
F01 & Tuition & My tuition is covered for my entire program. \\
F02 & Health & My college, department, or program offers me health benefits. \\
F03 & Life & I have no financial concerns about completing my degree. \\ \hline
\multicolumn{3}{l}{\textit{Social and scholarly exploration support} ($E$)} \\ \cline{1-3}
E01 & Socializing & The department hosts social activities (e.g., a welcome dinner, regular lunches) that are valuable in allowing me opportunities to share my thoughts and struggles with my peers, and discuss research areas. \\
E02 & Shared space & The department offered a space where students can build an academic and social community (e.g., student offices, rooms for tutoring, rooms for student leader organizations). \\
E03 & Accommodations & People in my department were supportive and caring about my accommodation needs when I first moved into town. \\
E04 & Peer mentor & I have or had a senior peer mentor that provided invaluable resources and inducted me into departmental and/or laboratory cultures. \\
E05 & Research match & I had or have support and flexibility from my department in finding my research interests. \\
E06 & Research exploration & I had or have the opportunity to rotate through different research labs without making a commitment in order to find my research match. \\
E07 & Research survey & I attend(ed) a research seminar surveying the areas of expertise within the department. \\
E08 & Research flexibility & My research mentor was very flexible with my research assignments when I was struggling with one or more courses. \\
E09 & Coursework support & Whenever I face(d) a challenge succeeding on coursework, someone from my department helped me overcome it. \\ \hline
\multicolumn{3}{l}{\textit{Mentoring and research experience} ($R$)} \\ \cline{1-3}
R01 & Research meetings & I have frequent meetings with my mentor to discuss on my research progress and any challenges I face. \\
R02 & Academic planning & My mentor(s) helped me selecting courses and develop my academic plans. \\
R03 & Informal meetings & I have informal meetings with my mentor(s) where I get assistance or support with any issues I face (for example, on issues such as life-work balance, develop social network, set future goals, access health care resources, etc.). \\
R04 & Academic integration & My mentor(s) helped me integrate into the program and the physics community. \\
R05 & Apprenticeship & My mentor(s) taught me what it means to be a research physicist and a scholar. \\
R06 & Meetings consistency & My research group meets at least once per week. \\
R07 & Journal discussions & In my research group meetings, we devote time in reading and discussing about the current state of knowledge in the field. \\
R08 & Regular feedback & I have regular meetings with my research mentor and receive feedback on a regular basis. \\
R09 & Project matching & The research project I am working on matches my research interests. \\
R10 & Presentations & I have presented or am planning to present my research at a group meeting or in a journal club. \\ \hline
\multicolumn{3}{l}{\textit{Professional and academic development} ($D$)} \\ \cline{1-3}
D01 & Academic assessment & In the beginning of my program, I took a precourse assessment that was designed to measure my incoming preparation. \\
D02 & Academic personalization & I was offered a personalized coursework plan in my graduate program. \\
D03 & Structured collaboration & The faculty, postdocs or experienced TAs lead guided group-work sessions to encourage students work collaboratively on concepts covered in core courses. \\
D04 & Networking & I attend mini-conferences where students from nearby universities can share research progress and learn networking skills. \\
D05 & Planning support & At the beginning of each semester, my faculty advisor(s) and I developed time-management plan that help me identify areas where my time could be used more effectively. \\
D06 & Time-management training & My department hosts a seminar that focuses on time-management skills. \\
E09 & Coursework support & Whenever I face(d) a challenge succeeding on coursework, someone from my department helped me overcome it. \\
D07 & Tutoring & My department makes tutoring available to graduate students. \\
D08 & Teaching training & I attend activities for graduate students that include trainings or professional development on best practices for effective teaching. \\
D09 & Postdoc training & I attend activities for graduate students that include trainings or professional development on the role of a postdoc. \\
D10 & Career training & I attend trainings that focus on how to maximize my chances of finding a career that is a good fit for my interests and skills. \\
D11 & Mentoring training & I attend training on learning about mentoring skills as future faculty or postdoc. \\
D12 & PI training & I attend training on organizing a research laboratory. \\
D13 & Networking training & I attend activities where I can learn about effective networking. \\ \hline \hline
\end{longtable}
\twocolumngrid
%%%%%%%%%%%%%%%%%%%%%%%%%%%%%%%%%%%%%%%%%%%%


%%%%%%%%%%%%%%%%%%%%%%%%%%%%%%%%%%%%%%%%%%%%
\end{document}